\language1
\magnification=1300
\vsize=23truecm
\hsize=15.6truecm
\tolerance=10000

\def\head #1 #2 { \headline={\vbox{ \line{\strut \TeX\hss #1--#2} \hrule}\hss} }
\def\ind {\noindent}
\def\tini {\mathop{\longrightarrow}}
\font\myfo=cmr8
\font\mybf=cmbx9
\font\myfon=cmr9

\myfon {
\footline{\hss\rm\quad\hss}
\centerline{\bf GENERALIZED MOYAL STRUCTURES IN PHASE-SPACE,}
\centerline{\bf KINETIC EQUATIONS AND THEIR CLASSICAL LIMIT}
\smallskip
\centerline{\bf I. GENERAL FORMALISM}
\vskip 1.0truecm
\centerline{CONSTANTINOS TZANAKIS  }
\centerline{\myfo University of Crete, 74100 Rethymnon, Crete, Greece }
\vskip 0.5truecm
\centerline{ALKIS P. GRECOS\footnote{$^*$}{Association Euratom, Etat Belge.}}
\centerline{\myfo Service de Physique Statistique et Plasmas}
\centerline{\myfo Universit\'{e} Libre de Bruxelles, Campus Plaine (CP 231)}
\centerline{\myfo 1050 Bruxelles, Belgium}
\vskip 1truecm
\ind ABSTRACT {\myfo
Generalised Wigner and Weyl transformations of quantum operators
are defined and their properties, as well as those of the algebraic
structure induced on the phase-space are studied. Using such transformations,
quantum linear evolution equations are given a phase-space representation.
In particular this is done for the general kinetic equation of the 
Lindblad type. The resulting expressions are better suited for the passage
to the classical limit and for a general comparison of classical and
quantum systems. In this context a preliminary discussion of a number
of problems of kinetic theory of open systems is given, whereas explicit
applications are made in the next paper of the series.}
\vfill{\eject}
\footline{\hss\rm\folio\hss}
\leftline{\bf 1. INTRODUCTION }
\bigskip
The formal similarity of the mathematical structures of classical and quantum
mechanics is perhaps best revealed in the context of statistical mechanics 
and kinetic theory. This is due to the formal similarity of their corresponding
starting points, the Liouville and von Neumann equations (Schr\"odinger
representation) or their formal adjoints (Heisenberg representation) which can
be put in the form:
$$\hskip 2truecm states\ \ \ \ \rho :\quad {\partial \rho  \over \partial t} =
 -iL\rho  \hskip 7.5truecm (1.1a)$$
$$\hskip 2truecm observables\ \ \ \ A:\ \ \ \ {\partial A \over \partial t} = i
LA \hskip 6.2truecm (1.1b)$$
$$\hskip 2.00664truecm L =  \cases {i\{H, \cdot\}&\ classical
\hskip 7.45truecm (1.2a) \cr {1 \over \hbar}[H, \cdot]& quantum
\hskip 7.45truecm (1.2b)} $$
Here $H$ is the system's Hamiltonian and $\{,\},[\ ,\ ]$ are respectively, the
Poisson bracket and operator commutator.
\par
In very broad terms, a fundamental problem in kinetic theory concerns the 
systematic derivation from dynamics, i.e.-from (1.1), of evolution equations for
the states or observables of the system and calculation of the corresponding
expectation values, by making well-defined assumptions concerning the system and
employing systematic approximation schemes to (1.1). Several such procedures 
exist, which at the level of the formalism, treat classical and quantum systems 
on an equal footing, most differences appearing at the level of applications
to specific systems (e.g. [1] and to a lesser extent [2]). However, the relation
between general results concerning classical and quantum systems is not always
clear. This is due to the fact that
\smallskip
\ind (i) there is no unique way of defining the quantum operator corresponding
to a classical phase-space function
\smallskip
\ind (ii) conversely, there is no unique way of mapping quantum operators to 
classical functions. Associated to this is the fact that there is a variety of
ways to pass to the classical limit of quantum mechanical expressions.
\par
The main reason for these difficulties is related to the fact that.
\smallskip
\ind (a) In (1.2), $L$ is defined on totally different spaces
\smallskip
\ind (b) $L$ in (1.2b), unlike $L$ in (1.2a) is defined through an associative
product. This is one of the reasons why quantum kinetic equations are sometimes
more easily derived than the corresponding classical ones.
\par
It is the purpose of the present series of papers to contribute to a unified
approach for the deduction from dynamics (i.e. from (1.1)) of kinetic
equations describing dissipative phenomena in classical and quantum systems,
 by using phase-space methods for both, so that the underlying space is the 
same. In this way, a clear comparison of the differences between classical and 
quantum systems imposed by the noncommutative mathematical structure 
characterizing the latter, will be more easily made. In particular, a clear
formulation of the passage to the classical limit of quantum mechanical
kinetic equations will be a more tractable problem. Moreover, conditions under
which this limit exists, or the reasons for which it may not exist, will be
more easily formulated.
\par
It is a well-known that phase-space methods in quantum theory originated through
the pionnering work of Weyl and Wigner, who gave respectively possible
solutions to the problems (i) and (ii) mentioned above, [3], [12]. Since the 
work of Moyal, [4], who showed the relation between the Weyl and Wigner 
transformations, many other possible transformations have been investigated,
particularly in the context of quantum optics [5], [25]. However it seems that
many researchers in this field adopted an attitude in which the following 
points dominate:
\smallskip
- to justify in one or another way the assertion that the Weyl transformation
and its inverse (the Wigner transformation) is the {\it only} possibility for
relating classical and quantum systems (see e.g. [6]).
\smallskip
- to develop quantum mechanics in phase-space {\it without} reference to the 
Hilbert state-space and some associated Banach algebra of linear operators. In
this connection the Moyal formulation of quantum mechanics 
(QM) is used, which amounts to introducing the structure of a 
nonabelian algebra for classical phase-space functions [6], [7], [8].
\par
However (a) the Wigner transformation presents severe problems if one wishes
to retain a probabilistic interpretation of QM in phase-space (specifically, it
does not conserve the positivity of the state).
\smallskip
(b) the Moyal formulation of QM is much more difficult to apply to specific 
problems. In fact simple systems studied in a most elegant way by 
conventional QM, turns out to require sophisticated mathematical
methods in the Moyal formulation [7].
\par
In our opinion a more "pragmatic" point of view is desirable: To accept that
both conventional QM and its phase-space formulation have their merits, exploit
their advantages as much as possible and use results known for classical 
systems to draw conclusions for quantum systems and vice versa. Adopting such
a point of view in the present series of papers, we are intented to 
contribute to a phase-space formulation of quantum kinetic equations and study
in this context some problems of kinetic theory which are described in the 
rest of this section.
\par 
After the introduction of the generalized Weyl and Wigner transformations in 
section 2,
their properties are further explored in section 3 and the associated 
generalized
Moyal algebras are studied. In section 4 the corresponding phase-space 
counterpart of quantum statistical evolution equations is given. Moreover 
we show that under quite general conditions, starting from (1.1), quantum 
kinetic theory of open systems in a projection operator language 
([9], [28], [21]) can be
developped in phase-space right from the beginning, using the algebra structure
induced on phase-space functions by generalized Wigner transformations. The
converse problem is studied in section 5, in which we obtain the phase-space
counterpart of a quantum kinetic equation of the Lindblad type, [24], which
as it is well-known, is the most general equation conserving the probabilistic
nature of the density matrix (the assumptions for its derivation need not be
discussed here!).
\par
In the second paper of this series (hereafter called paper II), we give 
specific applications of the general formalism to quantum harmonic oscillator
models, generalizing other approaches based on particular choices of 
operator orderings, and compare them with the
corresponding results for classical systems. Moreover we show that the classical
limit of the corresponding kinetic equations is independent of the generalized
Wigner transformation used, provided that the induced generalized Moyal
bracket is a deformation of the Poisson bracket.
\par
Finally in the third paper (hereafter called paper III) we will discuss certain
problems of kinetic theory of open systems, a brief account of which is the
following:
Various approaches in kinetic theory of open systems start from dynamics and
lead to kinetic equations of the Lindblad type for quantum systems ([22], [23],
 [28]) or its classical analogue ([9], [29]). A basic limitation of these 
approaches is that a certain Liouville operator\footnote{$^{(1)}$}{\myfo For 
open
systems it is the Liouville operator of the system when no interaction with its 
surroundings exists.} must have a discrete spectrum, an assumption particularly
restrictive for classical systems\footnote{$^{(2)}$}{\myfo Other formalisms not
exhibiting such a limitation lead to mathematically unacceptable or physically
incorrect results ([9] $\S$ 5, [10], [30], $\S \S$ 1, 2, [31] $\S
\S$ 1, 5).}. There are two possibilities out of this fundamental difficulty
\smallskip
(i) since for quantum systems (at least for finite ones) the above
restriction does not exist, we may consider the classical limit of the kinetic 
equation for the corresponding  quantum system. In this respect a phase-space
formulation of quantum kinetic theory is helpful.
\smallskip
(ii) To generalize the formalism relaxing this assumption. Such a formal 
generalization has been suggested for classical multiply periodic open
systems (i.e. integrable in action-angle variables) in [11].
\par
Unfortunately both possibilities face difficulties
\par
{\it For (i) }: In {\it particular} examples the classical limit does not exist
since the limiting equations contain divergent terms ([21] ch. 7). It is 
important
to decide whether or not this is a general feature. Or, to put it differently,
under what conditions the classical limit exists. In [11], section 4 it has been
conjectured that divergencies appear because of the structure of (1.2a),
particularly that $L$ there is a differential (hence unbounded) operator. A
unified phase-space formulation of kinetic theory, either classical or quantum 
is most suited for the examination of this problem (see also end of section 3
and footnote in section 4.1).
\par
{\it For (ii)}: The formalism in [11] has not yet been made mathematically 
rigorous. An {\it independent} test of its validity is to obtain kinetic 
equations for the corresponding quantum system and compare them with those of
the (generalized) classical formalism of [11]. Once again this pressuposes that
the (interesting per se) nontrivial old problem of finding the quantum analogues
of the classical multiply periodic systems has been answered definitely (see 
e.g. [19], [20], [13]). Phase-space methods may be proved fruitful in this 
context. In fact, one can consider the more general problem of giving an 
explicit expression of the generalized Weyl or Wigner transforms of canonically
or unitarily conjugate quantities respectively.
\par 
These and related questions will be examined in paper III in detail.
\vskip 2truecm 
\centerline{\bf 2. THE CORRESPONDENCE BETWEEN CLASSICAL}
\centerline{\bf PHASE-SPACE FUNCTIONS AND QUANTUM }
\centerline{\bf MECHANICAL OPERATORS}
\bigskip
In what follows we consider non-relativistic systems. Classsical
systems are assumed to be Hamiltonian and their phase-space $\Gamma$ is 
parametrized
by the canonical variables $(q,p)$. For the sake of simplicity the discussion
refers to an one-dimensional configuration space, but
all results are directly generalized in many dimensions. We use the following
notation:
\smallskip
\ind {\it phase-space functions}:\quad $A=A(q,p)$

\settabs 2 \columns
\+ {\it operators}:& {\it matrix elements}\cr
\+ position  $\hat q$ & $<x|\hat q|x'>=\delta (x-x')$\cr
\+ momentum  $\hat p$ & $<x|\hat p|x'>=-i\hbar {\partial \over \partial x}
\ \delta (x-x')$\cr
\+ general  $\hat A$ & $<x|\hat A|x'>\equiv A(x,x')$\cr
\+ density matrix  $\hat \rho $ & $<x|\hat \rho |x'>\equiv \rho (x,x')
\equiv <x,x'|\hat \rho  >$\cr
\smallskip
\ind Planck's constant \quad $\mu ={i\hbar \over 2}$
\medskip
\ind {\it Fourier transforms}:
$$\tilde A(\eta ,\xi )={1\over 2\pi }\int dqdp\ e^{-i(\eta q+\xi p)}A(q,p)$$
$$A(q,p)={1\over 2\pi }\int d\eta d\xi \ e^{i(\eta q+\xi p)}\tilde A(\eta,\xi)$$
Often we write
$$z=(q,p)\ ,\quad \hat z=(\hat q,\hat p)\ ,\quad \sigma =(\eta ,\xi )$$
Finally unless otherwise stated, we assume that $\hat q,\hat p$ have the whole
real line in their spectrum\footnote{$^{(3)}$}{\myfo It is possible to 
extend the formalism so that it includes the case when $\hat p $ has a point
spectrum. This is important in kinetic theory where often one first considers
confined systems and eventually pass to the thermodynamic limit of an infinite 
system having finite local properties. However we will not present the
corresponding calculations here because the resulting expressions are more 
complicated since in the limits of all integrations the size of the system
has to be taken into account.}.
The approach is formal but it will be seen that with no essential restrictions
we may suppose that all functions are generalized functions in either the
Schwarz space ${\cal D}'$ or the space of tempered distributions ${\cal I}'$
(see e.g. [26]).
\bigskip
\ind {\mybf 2.1 THE WEYL AND WIGNER TRANSFORMATIONS:}
\medskip
For any $A(q,p)$ its Weyl transform is the operator 
$$\hskip 1.5truecm \Omega _w:A(q,p)\tini \Omega_w(A)\equiv \hat A
\equiv {1\over 2\pi }\int d\sigma\ \tilde A(\sigma)e^{i\sigma\hat z}
\hskip 3.4truecm (2.1)$$
For any $\hat A$ its Wigner transform is the phase-space function 
$$\hskip 1truecm W:\hat A\tini W(\hat A)\equiv A(q,p)\equiv 2\int dt\ 
e^{{-pt\over \mu }}<q-t|\hat A|q+t>\hskip 1.6truecm (2.2)$$
A direct calculation using the identities
$$\hskip 2truecm e^{i(\eta \hat q+\xi\hat p)}=e^{i\eta \hat q\over 2}
e^{i\xi\hat p}e^{i\eta \hat q\over 2}=e^{\mu \eta \xi}e^{i\eta \hat q}
e^{i\xi\hat p} \hskip 5truecm (2.3)$$
$$\hskip 2truecm <x\mid e^{i(\eta {\hat q}+\xi\hat p)} \mid x'> = 
e^{i\eta (x-i\mu \xi)} \delta(x'-x+2i\mu \xi)\hskip 3truecm (2.4) $$
shows that
$$ W(\Omega _w(A)) = A,\ \ \ \ \ \ \ \ \ \ \Omega _w(W(\hat A)) = \hat A$$
so that $$W = \Omega_w^{-1}$$
Eq.(2.1) is defined so that $q^np^m$ corresponds to the Weyl-ordered operator
$$\hskip 2truecm \Omega _w(q^np^m) = \sum_{k=0}^n \pmatrix{n \cr k} \hat q^k 
\hat p^m \hat q^{n-k}\hskip 6truecm (2.5) $$
whereas (2.2) is defined so that for any $\hat \rho, \hat A$
$$\hskip 2.2truecm Tr(\hat \rho \hat A) = {1 \over 2\pi \hbar} \int \rho A
\ dpdq\hskip 7truecm (2.6a) $$
$$\hskip 2.1truecm \rho\equiv \Omega_w^{-1}(\hat \rho),\ \ \ \ \ \ \ \ \ \ A 
\equiv \Omega _w^{-1}(\hat A)\hskip 6truecm (2.6b) $$
In fact $ {1 \over 2\pi \hbar} \rho $ is the Wigner function associated with the
density matrix $\hat \rho$, hence quantum statistical expectations are obtained
 by
classical-type phase-space averages. However there are other orderings, distinct
from (2.5) and useful in applications, for which the inverse transformations 
satisfy (2.6). Before proceeding to their study we notice that if for any
phase-space functions $f,g$ we define the $*$-operation
$$\hskip 2truecm f * g = \Omega _w^{-1}(\Omega _w(f)\Omega _w(g))
\hskip 7.5truecm (2.7)$$
then $*$ endowes phase-space functions equipped with the ordinary structure of a
linear space $F(\Gamma )$, with the structure of a nonabelian, associative 
algebra. The commutator associated with the $*$-product is defined by
$$\hskip 2truecm {1 \over 2\mu }[f,g]_M = {1 \over 2\mu }(f*g-g*f)
\hskip 7truecm (2.8)$$
and for $\mu \rightarrow 0$ it gives the Poisson bracket $\{f,g\}$. Eq.(2.8)
defines the so-called Moyal product.
\bigskip
\ind {\mybf Remarks}: Because of (2.6), the Wigner transformation allows for a 
phase-space formulation of quantum statistical mechanics. In fact, appart from
(2.6), eq.(2.2) implies the following properties:

\ind If $\Omega _w^{-1}(\hat A) \equiv A_w(q,p) $ then
\smallskip
\ind (i) $\hat A = \hat A^+ \Leftrightarrow A_w = A_w^*$
\smallskip
\ind (ii) $Tr\hat A=1 \Leftrightarrow \int A_wdqdp = 2\pi \hbar $
\smallskip
\ind (iii) $\Omega _w^{-1}(\hat q^k) = q^k,\quad \ \Omega _w^{-1}(\hat p^k) 
= p^k $
\smallskip
\ind (iv) if $\hat A=\hat \rho=\mid \psi ><\psi \mid$ is a pure state then
$$ \int \rho_w(q,p)dp = 2\pi \hbar {\mid \psi(q) \mid}^2 $$
$$ \int \rho_w(q,p)dp = 2\pi  {\mid {\tilde \psi}({p \over \hbar}) \mid}^2 $$
Here $z^*$ is the complex conjugate of $z$.
\ind These properties will be considered in section 3 in a more general context.
We only notice that (2.6) together with (i), (ii) above, would imply that the
probabilistic interpretation of a density matrix $\hat \rho$ is retained in a 
phase-space formulation, provided that $ \rho_w \geq 0 $ whenever $\hat \rho 
\geq 0$.
However it is well-known that this is not true (see e.g. [2b] p.99, [4] p.116 
and
corrolary to proposition 3.4 below).
\vskip 3truecm
\ind {\mybf 2.2 GENERALIZED WEYL AND WIGNER TRANSFORMATIONS AND}
\ind {\mybf \ \ THE ASSOCIATED MOYAL STRUCTURE }
\medskip
A substantial and natural generalization of (2.1), which includes many useful
orderings of operators distinct from (2.5) are given by ([14])
\footnote{$^{(4)}$}
{\myfo It can be shown that any linear transformation of quantum operators to 
phase-space functions which is phase-space translation invariant is of the form
$(2.11')$. The calculations however will not be reproduced here. Compare with
[6b] in which Galilei invariance, conservation of hermiticity and normalization
of operators ensures that $(2.11')$ reduces to the Wigner transformation.}
$$\hskip 2truecm \Omega : A(q,p) \rightarrow \Omega (A) \equiv \hat A 
\equiv {1 \over 2\pi } \int d\sigma\ \Omega (\sigma )
{\tilde A}(\sigma ) e^{i\sigma \hat z}\hskip 2.7truecm (2.9)$$
For the weighting function $\Omega $ we assume that it is an entire complex 
analytic
function, with no zeros ([5a] p. 2166). The reason for this will become evident
below. If $A(q,p) = q^np^m$ then (2.9) gives an ordering different from (2.5).
 If
$$\hskip 2truecm \omega(z) = {1 \over 2\pi } \int d\sigma  e^{i\sigma z} 
{1 \over \Omega (\sigma )}\hskip 7.3truecm (2.10) $$
then the inverse $\Omega$ -transformation exists and is given by
$$\Omega^{-1}: \hat A \rightarrow A(q,p) = {1 \over \pi } \int dq'dp'dt
\omega(q-q',p-p') e^{-{p't \over \mu }} <q'-t \mid \hat A \mid q'+t>$$

\rightline {(2.11)} 
\ind The proof is straightforward, using (2.9), (2.10) and (2.3), (2.4). By 
(2.2) we
see that 
$$\hskip 1truecm \Omega ^{-1}(\hat A) = {1 \over 2\pi } \int dq'dp' 
\omega(q-q',p-p') A_w(q',p') ={1 \over 2\pi } \omega \odot A_w \hskip 1truecm 
(2.11')$$
where by $\odot$ we denote the convolution of two functions. 

\ind Consequently ${\widetilde {f\odot g}} = (2\pi {\widetilde {fg}})$, hence
$$\hskip 2truecm \Omega ^{-1}(\hat A) \equiv A(q,p) \Rightarrow {\tilde A}
(\sigma ) = {1 \over \Omega (\sigma )} {\tilde A}_w(\sigma )
\hskip 4.3truecm (2.12)$$
and by (2.2)
$$\hskip 2truecm {\tilde A}_w(\eta ,\xi ) = -2i\mu  \int dq\ e^{-i\eta q} 
<q+i\mu \xi \mid \hat A \mid q-i\mu \xi >\hskip 2truecm (2.2')$$
from (2.12) or directly from $(2.11')$, (2.10) we obtain that
$$ \Omega ^{-1}(\hat A) = {-i\mu  \over \pi  } \int dq'd\eta d\xi  
{e^{i[\eta (q-q')+\xi p]} \over \Omega (\eta ,\xi )}
<q'+i\mu \xi  \mid \hat A \mid q'-i\mu \xi >\ \ \ \ \ \ (2.11'') $$
Finally, in analogy with (2.7) we endow $F(\Gamma )$ with the structure of a 
nonabelian associative algebra, and consequently with a Lie algebra structure,
by defining for $f,g \in F(\Gamma )$
$$\hskip 2truecm f *_\Omega  g = \Omega ^{-1}(\Omega (f)\Omega (g))
\hskip 7.5truecm (2.13a)$$
$$\hskip 2truecm  {1 \over 2\mu  } [f,g] = {1 \over 2\mu } 
(f *_\Omega  g - g *_\Omega  f)\hskip 6.3truecm (2.13b)$$
Hereafter the Lie product in (2.13b) will be called the generalized Moyal
product and (2.9), (2.11) a generalized Weyl and Wigner transformation
respectively.
\bigskip
\ind {\mybf Remark}: We use the same symbol for the generalized
Weyl transformation $\Omega$ and for its kernel in (2.9). However since the
argument of the latter is always $\sigma=(\eta,\xi)$ etc, there is no 
confusion.  
\vskip 2truecm 
\leftline{\bf 3. THE $*_{\Omega}$ - ALGEBRAS AND THEIR PROPERTIES }
\bigskip
\ind {\mybf 3.1} In this section we will investigate the properties of (2.13).
 We
first give an explicit expression of (2.13). By (2.9) we have
$$ \Omega(f)\Omega(g) = {1 \over (2\pi )^2} \int d\sigma  d\sigma ' 
{\tilde f}(\sigma ) {\tilde g}(\sigma ') \Omega(\sigma ) \Omega(\sigma ') 
e^{(i\sigma {\hat z})}e^{(i\sigma '{\hat z})}$$
Using (2.3) we find that 
$$ e^{i\sigma {\hat z}}e^{i\sigma '{\hat z}}= e^{i{\hbar}\xi \eta '}
e^{i{\hbar}{\xi \eta +\xi '\eta ' \over 2}}e^{i{\hat q}(\eta +\eta ')}
e^{i{\hat p}(\xi +\xi ')} $$
$$ e^{i(\sigma +\sigma '){\hat z}} = e^{i{\hbar}{\xi \eta +\xi \eta '+\xi'
\eta +\xi '\eta ' \over 2}} e^{i{\hat q}(\eta +\eta ')}e^{i{\hat p}(\xi +\xi ')
} $$ 
hence
$$\hskip 2truecm e^{i\sigma {\hat z}} e^{i\sigma '{\hat z}} = 
e^{\mu(\sigma '{\land}\sigma)} e^{i(\sigma +\sigma '){\hat z}}
\hskip 6.9truecm (3.1)$$
where
$$ \sigma '{\land}\sigma  = (\eta ',\xi ') \land (\eta ,\xi ) = 
\eta'\xi  - \eta \xi ' $$
is the exterior product of $\sigma '$ and $\sigma$ . Substituting (3.1) in the
 expression for
$\Omega(f)\Omega(g)$ and using (2.13) we get
$$ (f *_{\Omega} g)(z) = {1 \over (2\pi )^2} \int d\sigma  d\sigma ' 
{\tilde f}(\sigma ) {\tilde g}(\sigma ') \Omega(\sigma ) \Omega(\sigma ') 
e^{\mu(\sigma ' \land \sigma )} \Omega^{-1}(e^{i(\sigma +\sigma '){\hat z}}) $$
But by (2.9) it is easily seen that 
$$\hskip 2truecm \Omega({e^{i\sigma z}}) = e^{i\sigma {\hat z}} \Omega(\sigma )
\hskip 8.7truecm (3.2) $$
hence
$$\ \ \ \ \ \ \ (f *_{\Omega} g)(z) = {1 \over (2\pi )^2} \int d\sigma  
d\sigma ' {\tilde f}(\sigma ) {\tilde g}(\sigma ')
{\Omega(\sigma )\Omega(\sigma ') \over \Omega(\sigma +\sigma ')} e^{\mu(\sigma'
 \land \sigma )}e^{i(\sigma +\sigma ')z}\ \ \ \ (3.3) $$
Another form of (3.3) can be found as follows: We define the commuting pairs 
of operators
$$ {\hat q} = -i {\partial \over \partial q},\ \ \ {\hat p} = -i {\partial 
\over \partial p},\ \ \ {\hat q'} = -i {\partial \over \partial q'},\ \ \ 
{\hat p'} = -i {\partial \over \partial p'}$$
having respectively eigenvalues $\eta ,\xi ,\eta ',\xi '$ and eigenprojections
 $F_\eta ,F_\xi ,F_{\eta '},F_{\xi '} $, where
$$ <q,p \mid F_\eta  \mid {\tilde q},{\tilde p}> = {e^{i\eta (q-{\tilde q})} 
\over 2\pi } \delta(p-{\tilde p})$$
$$ <q,p \mid F_\xi  \mid {\tilde q},{\tilde p}> = {e^{i\xi (p-{\tilde p})}
\over 2\pi } \delta(q-{\tilde q}) $$
and similarly for $F_{\eta '}, F_{\xi '}$. Then we readily find that for any 
functions
$f, g$.
$$ {e^{i\sigma z} \over (2\pi )^2} {\tilde f}(\sigma ) = (F_\eta F_\xi f)(z)\
 \ \ \ \ \ \ \ \ \ {e^{i\sigma 'z} \over (2\pi )^2} {\tilde g}(\sigma ) 
= (F_{\eta '}F_{\xi '}f)(z) $$
Therefore (3.3) can be rewritten as (c.f. [5a] p. 2190, eq(3.3))
$$\hskip 0.5truecm f *_\Omega g(z)=
(2\pi)^2\Omega(\hat z)\Omega(\hat z')
\Omega^{-1}(\hat z+\hat z')
e^{\mu(\hat z'\land \hat z)}f(z)g(z')\vert_{z=z'}\hskip 1.85truecm (3.3')$$
Eq. (3.3) readily implies that
$$\hskip 0.3truecm \widetilde{(f *_\Omega g)}(\sigma ) = 
{1 \over 2\pi} {1 \over \Omega(\sigma )} \int d\sigma ' 
e^{\mu(\sigma  \land \sigma ')}
\Omega(\sigma ') \Omega(\sigma -\sigma ') {\tilde f}(\sigma ') 
{\tilde g}(\sigma -\sigma ')\hskip 0.84truecm (3.4) $$
$$ {1 \over 2\mu}[f,g](z) = {1 \over (2\pi )^2}\int d\sigma  
d\sigma ' {\Omega(\sigma )\Omega(\sigma ') \over \Omega(\sigma +\sigma ')}
{\sinh\mu(\sigma '\land \sigma ) \over \mu} {\tilde f}
(\sigma ){\tilde g}(\sigma') e^{i(\sigma +\sigma ')z}\hskip 0.3truecm (3.5) $$
On the other hand, if we define the mapping
$$\hskip 2truecm f \rightarrow Uf:\quad Uf(z) = 
\int d\sigma\  \Omega(\sigma ){\tilde f}(\sigma ) 
e^{i\sigma z}\hskip 4.6truecm (3.6) $$
then we can easily show that
$$\hskip 2truecm U(f *_\Omega g) = Uf * Ug\hskip 8.5truecm (3.7) $$
thus proving
\bigskip
\ind {\mybf Proposition 3.1}: The $*$ and $*_\Omega$ products define isomorphic 
algebras on $F(\Gamma)$
\footnote{$^{(5)}$}{\myfo This is a special case of a more 
general result: If in (3.3) the kernel 
${\Omega(\sigma )\Omega(\sigma ') \over 
\Omega(\sigma +\sigma ')} e^{\mu(\sigma '\land \sigma )} $
is replaced by $ B(\sigma ,\sigma ') $ and we require this operation to 
be associative,
then (3.3) is the only possibility. This is closely related to the uniqueness
of the Moyal-algebra considered in [18]. These 
results will be presented elsewhere.}.
\bigskip
\ind {\mybf Proof}: Evidently (3.6) is linear and its inverse is
$$ f \rightarrow U^{-1}f:\quad\ U^{-1}f(z) = \int d\sigma  
{1 \over \Omega(\sigma )} {\tilde f}(\sigma ) e^{i\sigma z} $$
Moreover $\widetilde {Uf}(\sigma ) = 
{\tilde f}(\sigma )\Omega(\sigma )$, so that by (3.4) we have
$$ [U(f*_\Omega g)]^\sim(\sigma ) = 
{1 \over 2\pi} \int d\sigma ' e^{\mu(\sigma  \land \sigma ')}
\Omega(\sigma ')\Omega(\sigma -\sigma ') {\tilde f}(\sigma ')
 {\tilde g}(\sigma -\sigma ') $$
 where $ [U(f*_\Omega g)]^\sim(\sigma )$ is the Fourier transform of
 $ U(f*_\Omega g)$.
On the other hand by (3.4) with $\Omega=1$ we get the same expression for 
$ (Uf*Ug)^\sim(\sigma )$ Q.E.D.
\par
Eqs (2.8), (3.7) imply that
$$\hskip 2truecm {1 \over 2\mu} [f,g] = U^{-1} ({1 \over 2\mu} [Uf, Ug]_M )
\hskip 6.2truecm (3.8) $$
Moreover from (3.5) with $\Omega=1$ it is straightforward to show that
$$\hskip 2truecm \lim_{\mu \rightarrow 0} {1 \over 2\mu}[f,g]_M  = \{f, g\}
 \hskip 8.1truecm (3.9) $$
In general $\Omega(\sigma )$, hence $U$, depends on $\mu$. Let us then assume 
that $\lim\limits_{\mu \rightarrow 0}\Omega(\sigma )\equiv \Omega_0(\sigma )$ 
exists 
and is continuous and denote by $U_0$ the
corresponding transformation in (3.6). Then (3.8) implies 
$$\hskip 2truecm \lim_{\mu \rightarrow 0} {1 \over 2\mu}[f,g]= 
U^{-1}_0 \{U_0f,U_0g\}\hskip 6.5truecm (3.8')$$
If, in analogy with (3.9), we require that the l.h.s. of (3.8$'$) is 
$\{f,g\}$ then
$U_0$ must be an automorphism of the Poisson Lie algebra, i.e.
$$\hskip 2truecm U_0(\{f,g\}) = \{U_0f,U_0g\} \hskip 7.6truecm (3.10) $$
By (3.6) the r.h.s. of (3.10) is easily seen to be 
$$ \{U_0f,U_0g\}(z) = {1 \over (2\pi )^2} \int d\sigma  
d\sigma ' {\tilde f}(\sigma )\Omega_0(\sigma ) 
{\tilde g}(\sigma ')\Omega_0(\sigma ') 
e^{i(\sigma +\sigma ')z} (\sigma \land \sigma ') $$
whereas
$$\{\widetilde {f,g}\}(\tau) = 
{1 \over (2\pi )^2} \int dz' e^{-iz'\tau} \int d\sigma d\sigma '
e^{i(\sigma +\sigma ')z}{\tilde f}(\sigma ){\tilde g}(\sigma ') 
(\sigma  \land \sigma ') $$
so that we finally get from (3.6)
$$U_0\{f,g\}(z) = {1 \over (2\pi )^2} \int d\sigma d\sigma ' 
e^{i(\sigma +\sigma ')z} \Omega_0(\sigma +\sigma ')
{\tilde f}(\sigma ) {\tilde g}(\sigma ') (\sigma  \land \sigma ') $$
Therefore (3.10) is equivalent to
$$\Omega_0(\sigma +\sigma ') = \Omega_0(\sigma )\Omega_0(\sigma ') 
\Leftrightarrow \Omega_0(\sigma ) = c e^{k\sigma } $$
$ c,\ k\equiv (k_1,\ k_2) $ being constants.
If we further assume that $\Omega_0(q) = {\hat q}, \Omega_0(p) = {\hat p} $ 
then 
$\Omega_0(\sigma ) = 1$. Thus we have proved
\bigskip
\ind {\mybf Proposition 3.2}: If (i) $\lim\limits_{\mu \rightarrow 0} 
{1 \over 2\mu} [f,g] = \{f,g\}$

\ind (ii) $\lim\limits_{\mu \rightarrow 0} \Omega(\sigma ) = \Omega_0(\sigma )
 $ is continuous and $(q,p) \rightarrow$
$({\hat q},{\hat p})$ via ${\Omega_0}$ then $\Omega_0(\sigma ) = 1$. If in 
addition $\Omega$ is
$\mu$-independent then the only $*_\Omega$-product, the associated Lie bracket
 of
which tends to the Poisson bracket as $\mu \rightarrow 0$ is the Moyal product.
\bigskip
\ind {\mybf 3.2}: As already stated in section 2, the main motivation for
introducing the Wigner transform is eq(2.6). Here we generalize this equation
for the $*_\Omega$-product:
\bigskip
\ind{\mybf Proposition 3.3}: For any $f, g \in F(\Gamma ) $
$$\hskip 2truecm Tr(\Omega(f)\Omega(g)) = {\Omega(0) \over 2\pi  \hbar} 
\int f*_\Omega g\ dz\hskip 5.7truecm (3.11)$$
\bigskip
\ind {\mybf Proof }: By (2.9) we have
$$ Tr(\Omega(f)\Omega(g)) = {1 \over (2\pi )^2} \int d
\sigma  d\sigma ' {\tilde f}(\sigma ){\tilde g}(\sigma ')
\Omega(\sigma )\Omega(\sigma ') Tr(e^{\sigma {\hat z}} e^{\sigma '{\hat z}}) $$
From (3.1), (2.4) we have 
$$Tr (e^{\sigma {\hat z}} 
e^{\sigma '{\hat z}}) = e^{\mu(\sigma ' 
\land \sigma)}Tr (e^{i(\sigma +\sigma '){\hat z}}) 
= e^{\mu(\sigma ' \land \sigma )}
{2\pi  \over \hbar } \delta(\sigma +\sigma ') $$
hence
$$\hskip 2truecm Tr(\Omega(f)\Omega(g)) = 
{1 \over 2\pi  \hbar} \int d\sigma  {\tilde f}(\sigma ) {\tilde g}(-\sigma )
\Omega(\sigma )\Omega(-\sigma )\hskip 2.6truecm (3.12) $$
On the other hand (3.3) implies
$$\hskip 2truecm \int f *_\Omega g\ dz = 
\int d\sigma {\tilde f}(\sigma )
{\tilde g}(-\sigma ) {\Omega(\sigma )\Omega(-\sigma ) 
\over \Omega(0)}\hskip 4truecm  (3.13) $$ 
and consequently (3.11) follows Q.E.D.
\bigskip
\ind {\mybf Remarks}: (i) Eq (3.11) immediately implies that
$$ Tr(\Omega(f)) = {1 \over 2\pi  \hbar } \int f dz $$
\hskip 2.1truecm (ii) $ \int f *_{\Omega} g\ dz = \int g *_{\Omega} f dz $
\bigskip
\ind {\mybf Corrolary}: A neccesary and sufficient condition for
$$\hskip 2truecm Tr(\Omega(f)\Omega(g)) = 
{1 \over 2\pi \hbar} \int f g\ dz\hskip 6.2truecm (3.14) $$
is
$$\hskip 2truecm \Omega(\sigma )\Omega(-\sigma ) = 
\Omega(0)\hskip 8.54truecm (3.15) $$
\ind{\mybf Proof}: Since $\int f g dz = 
\int {\tilde f}(\sigma ){\tilde g}(-\sigma ) d\sigma $,  
(3.15) follows from (3.12)-(3.14) Q.E.D.
\ind Further properties of the $*_\Omega$-structures are contained in 
the following
\bigskip
\ind {\mybf Proposition 3.4}: (a) for any $f \in F(\Gamma)$, the requirement 
that $f$
is real if and only if $\Omega(f)$ is hermitian, is equivalent to
$$\hskip 2truecm \Omega(\sigma ) = \Omega^*(-\sigma )\hskip 9.3truecm (3.16) $$
In that case 
$$\hskip 2truecm ( f *_\Omega g)^* = g^* *_\Omega f^* \hskip 8.3truecm  (3.16')
 $$
\ind (b) The conditions 
$$\hskip 2truecm \Omega(0,\xi ) = \Omega(\eta ,0) = 1 \ \ \ \ \ \forall \eta ,
\xi  \in {\cal R}\hskip 5.7truecm (3.17) $$
are equivalent to either of the following 
\medskip
-$\  \Omega(f(q)) = f({\hat q}),\ \ \ \ \  \Omega(g(p)) = g({\hat p})\ \ \ \ \
  \forall f,g \in F(\Gamma )\hskip 3.2truecm  (3.18) $
\medskip
- If ${\hat \rho}=\mid \Psi ><\Psi \mid $   is a pure state then
$$\hskip 2truecm \int \Omega^{-1}({\hat\rho})dp=-4i\mu \pi {\mid \Psi (q)\mid }
^2\hskip 6.2truecm (3.19a) $$
$$\hskip 2truecm \int \Omega^{-1}({\hat\rho}) dq = 2\pi  {\mid {\tilde \Psi }
 ({p \over -2i\mu }) \mid }^2\hskip 6.1truecm  (3.19b) $$
\ind {\mybf Proof}: Eq (3.16) follows from ${\tilde A}^{*}(\sigma ) 
= ({\tilde A}(-\sigma ))^{*}$, whereas (3.18) is an immediate consequence of 
(2.9). For the 
proof of (3.19) we proceed as follows:

\ind From $(2.11'')$ we readily get 
$$ \Omega^{-1}({\hat \rho})=- { i\mu  \over \pi  } \int dq' d\eta  d\xi\  
{e^{i[\eta (q-q')+\xi p]}\over \Omega(\eta ,\xi )} \Psi (q'+i\mu \xi ) 
\Psi ^*(q'-i\mu \xi )\hskip 1.4truecm (3.20) $$
hence
$$ \int \Omega^{-1}({\hat \rho}) dp = -2i\mu  \int dq' d\eta\ 
{ e^{i\eta (q-q')}\over \Omega(\eta ,0)}
{\mid \Psi (q') \mid}^2 $$
hence $\Omega(\eta ,0) = 1 $ is equivalent to (3.19a). Eq (3.20) implies
$$ \int \Omega^{-1}({\hat \rho}) dq =  -2i\mu  \int dq'd\xi\  
{e^{i\xi p} \over \Omega(0,\xi )} \Psi (q'+i\mu \xi )
\Psi^* (q-i\mu \xi ) $$
But since $\Psi (q) = {1 \over \sqrt{2\pi }} \int dk 
e^{ikq}{\tilde \Psi }(k)$ substitution
gives after some reductions, that 
$$ \int \Omega^{-1}({\hat \rho}) dq = -4i\mu \pi  \int dkd\xi\  
{e^{i\xi (p+2i\mu k)} \over \Omega(0,\xi )} 
{\mid {\tilde \Psi }(k) \mid}^2 $$
hence $\Omega(0,\xi ) = 1$ is equivalent to (3.19b) Q.E.D.
\bigskip
\ind {\mybf Remark}: For any operator ${\hat \rho}$, (3.17) is equivalent to
$$\hskip 2truecm  \int \Omega^{-1}({\hat \rho}) dp = -4i\mu \pi  
<q\mid {\hat \rho}\mid q> \hskip 5truecm (3.19'a) $$
$$\hskip 2truecm \int \Omega^{-1}({\hat \rho}) dq = 2\pi  
<{p \over -2i\mu }\mid {\hat \rho}\mid {p \over -2i\mu }>\hskip 4truecm  
(3.19'b) $$
where the r.h.s. in $(3.19'b)$ is in momentum representation.
\bigskip
\ind {\mybf Corrolary}: Conservation of hermiticity and (3.14) implies that
the generalized Wigner transform {\it does not conserve} the positivity of
the density matrix.
\bigskip
\ind {\mybf Proof}: In the present notation and for any two phase-space 
functions
we have by (2.12)
$$ \int dz A^{*}(z)B(z) =$$
$$ {1\over 2\pi } \int ({\tilde A}(\sigma ))^* {\tilde B}(\sigma )d\sigma  =
{1 \over 2\pi } \int d\sigma  {1 \over \Omega^{*}(\sigma )\Omega(\sigma ) }
({\tilde A}_w(\sigma ))^* {\tilde B}_w(\sigma ) $$
\ind If $A,B$ correspond to pure states, i.e.
$$\Omega(A) = \mid \Psi ><\Psi  \mid,\ \ \ \ \ \Omega(B) = \mid \Phi ><\Phi 
 \mid $$
then by $(2.11')$ an elementary calculation gives
$$ \int dz A^{*}(z)B(z) =$$
$${{\hbar}^2 \over (2\pi )^3 }\int d\eta d\xi dydx\ \ 
{e^{i\eta (y-x)} e^{-i\hbar \xi \eta } \over \Omega(\eta ,\xi )\Omega^{*}(\eta
 ,\xi )} (\Phi (x)\Psi ^{*}(x))
(\Phi (y)\Psi ^{*}(y))^{*} $$
Therefore by (3.14), (3.16) we get
$$ \int dz A(z)B(z) = {\hbar \over (2\pi )^2\Omega(0)} 
{\mid <\Phi  \mid \Psi > \mid}^2 $$
since (3.16) implies that $A, B$ are real.
\ind Therefore for a given $\Psi$, we can always choose a $\Phi$ orthogonal to
 it, and
consequently either $A$ or $B$ become negative somewhere, since we assume that 
$\Phi , \Psi$  are not identically zero Q.E.D.
\bigskip
\ind {\mybf 3.3}: Irrespective of any considerations concerning the 
correspondence
between classical and quantum systems, there is a very simple way to define a
structure of a nonabelian algebra on $F(\Gamma )$. The motivation comes from
considerations conserning the relation between stochastic mechanics 
(specifically the It${\hat o}$ calculus) and noncommutative geometry [15], [30]:
We shall show below that this structure is induced on $F(\Gamma )$ by a 
particular $\Omega$-transformation.  We define a $*_\lambda $-product as 
follows:
$$\hskip 2truecm q *_\lambda  p = qp + (\lambda +\nu )\hskip 8truecm (3.21a) $$
$$\hskip 2truecm  p *_\lambda  q = qp + (\lambda -\nu )\ \ \hskip 3truecm 
\lambda ,\nu  \in {\cal C}\hskip 3.18truecm (3.21b) $$
and we require:
\par
(a) $*_\lambda $ is associative and distributive with respect to vector-space
addition in $F(\Gamma )$.
\par
(b) For any $f(q), g(q), h(p), \sigma (p)$ we have 
$$\hskip 2truecm  f(q) *_\lambda  (g(q) h(p)) = (f(q) *_\lambda  h(p)) 
g(q)\hskip 4.9truecm  (3.22a) $$
$$\hskip 2truecm  \sigma (p) *_\lambda  (g(q) h(p)) 
= (\sigma (p) *_\lambda  g(q)) h(p)\hskip 4.9truecm  (3.22b) $$
and the same relations hold for multiplication by 
$f,\sigma$ from the right. Then it
can be shown that for $C^{\infty}$-functions $f, g$ ([15]):
$$(f *_\lambda  g)(q,p) = 
e^{[\lambda ({\partial 
\over \partial q}
{\partial \over 
\partial p'} + {\partial \over \partial p}{\partial \over \partial q'})
+\nu  ({\partial \over \partial q}
{\partial \over \partial p'} - 
{\partial \over \partial p}
{\partial \over \partial q'})]}f(z)g(z'){\mid}_{z=z'}\hskip 2.1truecm (3.23) $$
or in the notation of $(3.3')$
$$ (f *_\lambda  g) =
e^{[-\lambda ({\hat q}{\hat p}'
+{\hat q}'{\hat p})+\nu ({\hat q}'{\hat p} - 
{\hat q}{\hat p}')]}f(z)g(z'){\mid}_{z=z'} $$
Comparison with (3.3$'$) shows that the $*_\lambda $-product is induced by a 
generalized Wigner transformation, provided that 
$\nu=\mu ,\ \ \ \Omega(\eta ,\xi ) = e^{\lambda \xi \eta } $ .
Conversely we will now show.
\bigskip
\ind {\mybf Proposition 3.5}: A necessary and sufficient condition for a
$*_\Omega$-product to satisfy (3.22) and $\Omega:
 (q,p) \rightarrow ({\hat q},{\hat p})$,
is that $\Omega(\eta ,\xi ) =e^{\lambda \eta \xi }$
\bigskip
\ind {\mybf Proof}: For any $f(q),\ g(q),\ h(p)$ we have
$$ {\tilde f}(\eta ,\xi) = \sqrt{2\pi } {\tilde f}(\eta ) \delta (\xi ) $$
$$\widetilde {gh}(\eta ,\xi ) = {\tilde g}(\eta ){\tilde h}(\xi ) $$
$$ {\tilde h}(\eta ,\xi ) = \sqrt{2\pi } {\tilde h}(\xi ) \delta (\eta ) $$
hence by (3.3)
$$ (f *_\Omega gh)(q,p) =$$
$$ {1 \over (2\pi )^{({3\over 2})}} 
\int d\eta d\xi  {\tilde f}(\eta ) 
{\tilde h}(\xi ) e^{-\mu \eta \xi } 
e^{i[q\eta +\xi p]}\Omega(\eta ,0) 
\int d\eta 'e^{i\eta 'q} {\Omega(\eta ',\xi ) 
\over \Omega(\eta +\eta ',\xi )} {\tilde g}(\eta ') $$
On the other hand
$$ (f *_\Omega  h)(q,p) = 
{1 \over 2\pi } \int d\eta d\xi '\ {\tilde f}(\eta ) {\tilde h}(\xi ') 
e^{-\mu \eta \xi '} {\Omega(\eta ,0) 
\Omega(0,\xi ') \over \Omega(\eta ,\xi ')} e^{i(q\eta +\xi 'p)} $$
so that
$$(g(f *_\Omega  h))(q,p) =$$
$${1 \over (2\pi )^{({3\over 2})}} 
\int d\eta d\xi\ {\tilde f}(\eta ) {\tilde h}(\xi )
e^{-\mu \eta \xi } e^{i[q\eta +\xi p]} \Omega(\eta ,0) 
\int d\eta ' e^{i\eta 'q} {\Omega(0,\xi ) \over 
\Omega(\eta ,\xi )} {\tilde g}(\eta ') $$
Therefore, comparison shows that for (3.22a) to hold it is necessary and
sufficient that 
$$ \Omega(\eta ',\xi )\Omega(\eta ,\xi ) = \Omega(0,\xi )
\Omega(\eta +\eta ',\xi ) $$
Logarithmic differentiation shows that the solution is
$$ \Omega(\eta ,\xi ) = \Omega (0,\xi ) e^{\lambda (\xi )\eta } $$
In exactly the same way we find that (3.22b) is equivalent to
$$ \Omega(0,\xi')\Omega(\eta ,\xi ) = \Omega(\eta ,0)\Omega(\eta ,\xi +\xi ') $$
so that 
$$ \Omega(0,\xi ')\Omega(0,\xi ) 
e^{[\lambda (\xi )+\lambda (\xi ')]\eta } = 
\Omega(0,0)\Omega(0,\xi +\xi ')e^{[\lambda (\xi +\xi ')+\lambda (0)]\eta } $$
For $\eta =0$ and by logarithmic differentiation we once again show that 
$$\Omega(0,\xi ) = \Omega(0,0)e^{a\xi }$$
hence $\lambda (\xi )+\lambda (\xi ') = 
\lambda (\xi +\xi ')+\lambda (0)$, from which we obtain 
$\lambda (\xi ) = \lambda \xi +b$. Therefore
$$\Omega(\eta ,\xi ) = \Omega(0,0) e^{\lambda \eta \xi } e^{a\xi +b\eta }$$
Requiring that $\Omega(q) = {\hat q},\ \ \Omega(p) = 
{\hat p} $ implies $\Omega(0,0) = 1,\ a=b=0 $
Q.E.D.
\smallskip
The dependence of $*_\lambda $ on the two parameters $\lambda ,\mu $ gives 
the opportunity to
consider in more detail the problems mentioned at the end of section 1:
\ind Since in the classical limit, the von-Neumann operator often has a
continuous spectrum (e.g. think of a free particle in a box and that
${1 \over \hbar}[{\hat p}^2,\cdot] \rightarrow p{\partial \over \partial q}$
as $\hbar \rightarrow 0^+$), it is important to be able to keep control of the
discreteness of the spectrum, depending e.g. on $\lambda$, and quantum 
corrections,
depending on $\mu$. In this way divergencies appearing in the derivation of 
classical kinetic equations from dynamics discussed in section 1 can be better
understood. This will be discussed in paper III (c.f. footnote in section 4, end
of section 4 and [11] setion 4). Here we only notice that for $\mu =0$,  
$*_\lambda $ is a
commutative product, different from the ordinary one. Moreover, for the 
sake of completeness we remark that from (3.4), (3.3) we can show that if 
$\Omega_\lambda $
is the transformation corresponding to $\Omega(\eta ,\xi ) = 
e^{\lambda \eta \xi }$, then 
$$\Omega_\lambda (q^n\ p^m) = \sum_{l=0}^m (-i\lambda )^l \pmatrix{m \cr l} 
\pmatrix{n \cr l}
l! \Omega_w (q^{n-l} p^{m-l}) $$
where $\Omega_w$ is the Weyl transformation, eq.(2.5) (c.f. eq (4.4) in [15a]). 
Moreover 
$$ (f *_\lambda  g)(q,p) =$$
$${4 \over \pi ^2} e^{2i\theta } 
\int dydy'd\rho d\rho ' 
\{ f(q+y+y',p-2ia(\rho -\rho ')) g(q+y-y',p-2ia(\rho +\rho '))$$
$$ e^{2i[(y\rho +y'\rho ')(1+e^{2i\theta })+
(y'\rho +y\rho ')(1-e^{2i\theta })]}\} $$
where
$$ a=\lambda +\mu ,\ \ \ \ \ \theta =\arctan(-i {\mu \over \lambda })$$
For $\theta = {\pi  \over 2},\ \ \mid a \mid = 
{\hbar \over 2}$ we get eq.(3.23) of [6a] for
the $*$-product eq.(2.7).
\vskip 2truecm 
\leftline{\bf 4. QUANTUM EVOLUTION EQUATIONS IN PHASE-SPACE}
\leftline{\bf \hskip 0.64truecm REPRESENTATION}
\bigskip
\ind {\bf 4.1}: As already stated in section 1, one of the basic
motivations for introducing the Wigner transform is the search for 
a phase-space formulation of quantum statistical mechanics. Here we
explicit such a possibility.
\par
Suppose that the quantum statistical state $\hat \rho$ satisfies the 
{\it linear} evolution equation
$$\hskip 3truecm i{\partial \hat \rho \over \partial t}=\hat \Phi \hat \rho 
\hskip 9.7truecm (4.1)$$
where $\hat \Phi$ is a superoperator acting on the state space of the
quantum system. In position representation (4.1) becomes (cf. the notation
at the beginning of section 2)
$$\hskip 1truecm i{\partial \over \partial t}<x,y|\hat \rho>=
\int dx'dy'<x,y|\hat \Phi|x',y'> <x',y'|\hat \rho>\hskip 2.6truecm (4.1')$$
Putting $\Omega^{-1}(\hat \rho)\equiv \rho(q,p)$ we try to rewrite $(4.1')$
in the equivalent form
$$\hskip 1truecm i{\partial \over \partial t}\rho(q,p)=
\int dq'dp'<q,p|\Phi|q',p'> \rho(q',p')\hskip 4.5truecm  (4.2)$$
and express the generator in phase-space $\Phi$, in terms of $\hat \Phi$
and $\Omega$. Putting $x=q'-t,\ y=q'+t$ in (4.1$'$) and using (2.11) we
readily find
$$i{\partial \over \partial t}\rho(q,p)=$$
$${1\over \pi}\int dq'dp'dtdx'dy'\omega(q-q',p-p')
e^{-p't\over \mu}<q'-t,q'+t|\hat \Phi|x',y'><x',y'|\hat \rho>$$
Making the substitutions $x'=q''-s,\ y'=q''+s$, using (2.4) in (2.9) to
express $<x|\hat \rho|y>$ in terms of $\rho$ and substituting above we finally
get
$$i{\partial \over \partial t}\rho(q,p)
={1\over 2\pi ^2}\int dq'dp'dtdq''dsd\eta d\xi\ 
\omega(q-q',p-p')e^{-p't\over \mu}$$
$$<q'-t,q'+t|\hat \Phi|q''-s,q''+s>\tilde 
\rho(\eta,\xi)\Omega(\eta,\xi)e^{i\eta q''}\delta(s-{\hbar \xi\over 2})$$
Writing $\delta(s-{\hbar \xi\over 2})=
{1\over 2\pi}\int dp''e^{ip''({\hbar \xi\over 2}-s)}$, 
changing from $p''$ to ${hp''\over 2}$ and using (2.12), (2.10), 
we find
$${\partial \over \partial t}\rho(q,p)=
{1\over \pi^2\hbar}\int dq'dq''dp'dp''d\tilde qd\tilde pdsdt\ 
e^{-{p't\over \mu}} \omega(q-q',p-p')$$
$$\hskip 1truecm <q'-t,q'+t|\hat \Phi|q''-s,q''+s>\omega
(q''-\tilde q,p''-\tilde p)e^{{sp''\over \mu}}\rho(\tilde q,\tilde p)
\hskip 2truecm (4.3)$$
Comparison with (4.2) gives the desired expression for $\Phi$:
$$<q,p|\Phi|q',p'> = {1\over \pi^2 \hbar} \int d\tilde qd\tilde 
pd\tilde q'd\tilde p'dsdt\ e^{-{\tilde pt\over \mu}}\hskip 6.1truecm$$
$$ \omega(q-\tilde q,p-\tilde p)<\tilde q-t,
\tilde q+t|\hat \Phi|\tilde q'-s,\tilde q'+s>\omega(\tilde q'-q',
\tilde p'-p')e^{s\tilde p'\over \mu}\hskip 1.3truecm (4.4)$$
This expression is completely general for any {\it linear} $\hat \Phi$, hence
more difficult to apply to spesific examples. We may notice however that in
actual applications, $\hat \Phi$ {\it is usually constructed by using only the
algebra structure of quantum operators or operations that can be considered as
limiting cases of algebraic operations} (e.g. integration of an operator
depending on some parameter, with respect to this parameter). Then it is clear
that since $\Omega^{-1}$ is an algebra homomorphism (c.f.(2.13)) and if
$\hat \Phi = F(\hat A,\hat B,...)$ then from (4.1) we obtain that
$$\hskip 2truecm i{\partial f \over \partial t}=
\Phi f\equiv(\Omega^{-1}(\hat \Phi))f\hskip 7.5truecm  (4.4'a)$$
$$\hskip 2truecm \Omega^{-1}(\hat \Phi)
\equiv F(\Omega^{-1}(\hat A),\Omega^{-1}(\hat B),...)\hskip 5.6truecm (4.4'b) $$
Here {\it F is expressed as a function of its arguments, by using the 
$*_{\Omega}$-product.}
Notice also that by (3.11), $(4.4')$, expectation values evolve according to
$$\hskip 2truecm i{\partial \over \partial t}\int(A*_{\Omega}\rho)\ dz
=\int A*_{\Omega}(\Phi\rho)\ dz\hskip 5.5truecm (4.5) $$
\par
Moreover, $(4.4')$ shows that at least formally $\Phi,\hat \Phi$ have the same
spectrum and their eigenprojections are $\Omega$-transforms of each other. In
this way not only the classical limit of a quantum kinetic equation can be 
studied but also the spectral representation of its limit, if it exists, can be
obtained. This approach can be used in two different ways:
\medskip
(i) Either we start from dynamics, (1.1), obtain a quantum kinetic equation 
(4.1)
by some systematic procedure and then take its $\Omega$-transform to get its
phase-space representation 

or (ii) (1.1) is {\it directly} $\Omega$-transformed and by the {\it same} 
procedure we obtain a quantum kinetic equation in phase-space.
\par
Clearly the two ways must be compatible, or schematically the following diagram
must be commutative.
$${\partial\hat\rho\over\partial t}=-i\hat L\hat \rho\hskip 2truecm 
\longrightarrow\hskip 2truecm {\partial\hat\rho\over \partial t}=
-i{\hat\Phi\hat\rho}$$
$$\Omega ^{-1}\downarrow\hskip 6truecm \downarrow \Omega ^{-1}$$
$${\partial \rho\over\partial t }=-iL^{\Omega}\rho\hskip 1.8truecm 
\longrightarrow\hskip 2truecm {\partial\rho\over\partial t}=-i\Phi\rho$$
where $L^{\Omega}=\Omega^{-1}(\hat L)=[\Omega^{-1}(\hat H),\cdot]$
(c.f.(2.13)).

By the {\it first} method, which is easier to apply, a quantum kinetic formalism
in phase-space can be compared with a corresponding one for classical systems 
and be taken as a starting point for taking the classical limit of quantum 
kinetic equations. This is considered in the next section and paper II. Below we
will apply the {\it second} method in the context of the general formalism for
 open
systems developped in [1b], [28], [9]. Specifically we will show that, using
phase-space functions it is possible to write down meaningful quantum kinetic
equations, formally identical to their classical analogues, in which the 
ordinary product of functions is replaced by the   $*_{\Omega}$-product
\footnote{$^{(6)}$}
{\myfo This may be helpful in examining the conditions under which the formalism
gives well-defined results in the limit in which the $*_{\Omega}$-product and/or
the assosiated bracket reduce to the ordinary product and the Poisson bracket
respectively. This is closely related to the problems addressed in paper III,
briefly discussed at the end of section 1 (see the comments following
proposition 3.5 and end of this section).}.
\par
{\it Compatibility of the two methods mentioned above} follows then from the 
results of the next section, namely that in the corresponding quantum formalism
the resulting quantum equations, when $\Omega$-transformed, are identical to 
their classical counterparts expressed with the aid of the $*_{\Omega}$-product.
\medskip
{\bf 4.2} As mentioned above, we consider an open system $\Sigma$, interacting
 with a much
larger one, the reservoir $R$, which is originally in an equilibrium state
$\rho_R$. We suppose that the Liouville operator (1.2) is
$$\hskip 2truecm L=L_{\Sigma}+L_R+\lambda L_I\ \equiv L _0 +\lambda L_I 
\hskip 5.7truecm  (4.6a)$$
$$\hskip 2truecm H=H_{\Sigma}+H_R+\lambda H_I\hskip 8.5truecm  (4.6b)$$
with $L_{\Sigma}$ defined by $H_{\Sigma}$ and similarly for $L_R,L_I$, and where
$L_{\Sigma}, L_R$ depend only on the phases of $\Sigma, R$ respectively and
$L_I$ is an interaction term, $\lambda$ being the coupling parameter. It is 
possible to develop a general formalism for the time evolution of the state
of $\Sigma$, assuming that $R$, being much larger than $\Sigma$, practically
remains in the equilibrium state $\rho_R$ [9]. The formalism uses 
projection-operator
methods and leads to explicit results for spesific systems (see e.g. [16]).

In particular for weak-coupling (i.e. omitting terms of order higher than 
$\lambda ^2$) the formalism has been studied in [9]. {\it Below we show that the
same formalism can be developped for quantum systems, starting from the
$\Omega$-transform of the von Neumann equation}. Mathematically speaking this is
identical to expressing this formalism for {\it classical} systems with the
$*_{\Omega}$-product replacing the ordinary product of functions.
\par
It is clear that {\it the only points of the formalism needing special 
consideration are those involving products of functions}.

A careful analysis of [9] shows that {\it these are:}
\medskip
(i) The definition of the projection $P$ on the state of ${\Sigma}$
$$\hskip 1.7truecm P\rho(z_{\Sigma},z_R,t)\equiv \rho_R(z_R)
\int \rho(z_{\Sigma},z_R,t)\ dz_R
\equiv \rho_R(z_R)f_{\Sigma}(z_{\Sigma},t)\hskip 1truecm  (4.7)$$
where $z_{\Sigma}=(q_{\Sigma},p_{\Sigma}),\ z_R=(q_R,p_R) $ are the phases of
${\Sigma},R$ respectively. From this we have for its adjoint
$$\hskip 1.8truecm P^+A(z_{\Sigma},z_R)=
\int\rho_R(z_R)A(z_{\Sigma},z_R)\ dz_R\hskip 4.8truecm (4.7')$$
\medskip
(ii) The basic properties of $P$, namely
$$\hskip 2truecm PL_{\Sigma}=L_{\Sigma}P\hskip 9.8truecm  (4.8a)$$
$$\hskip 2truecm PL_R=L_RP=0 \hskip 9truecm  (4.8b)$$
\medskip
(iii) $L,\ \ L_{\Sigma},\ \ L_R,\ \ L_I$ are formally self-adjoint.
\medskip
If (4.7), $(4.7')$, (4.8) can be verified for a $*_{\Omega}$-product then the
{\it whole formalism is valid and any existing kinetic equation in that 
formalism is obtained by rewritting it using the $*_{\Omega}$-product.}

In the notation of section 2, with subscripts to distinguish between ${\Sigma}$
and R, we define
$$\hskip 1.5truecm P\rho(z_{\Sigma},z_R,t)
\equiv\rho_R(z_R)*_{\Omega}\int\rho\ dz_R
\equiv\rho_R(z_R))*_{\Omega}f_{\Sigma}(z_{\Sigma},t)\hskip 2truecm (4.9) $$
where $\rho_R$ is such that 
$$\hskip 2truecm [H_R,\rho_R]=0\hskip 9.9truecm  (4.10)$$
In general $\rho_R*_{\Omega}f_{\Sigma}\neq\rho_Rf_{\Sigma}$ but it is 
{\it reasonable to require} that if $A_{\Sigma}=A_{\Sigma}(z_{\Sigma}),\ \ 
A_R=A_R(z_R)$ and if $\Omega(A_{\Sigma})\equiv
\hat A_{\Sigma},\ \ \Omega(A_R)\equiv\hat A_R$
then $\Omega(A_{\Sigma} A_R)=\hat A_{\Sigma}\hat A_R$. Then we readily find that
$$\hskip 2truecm \Omega(\sigma_{\Sigma},\sigma_R)=\Omega(0,\sigma_R)
\Omega(\sigma_{\Sigma},0)
\equiv\Omega(\sigma_R)\Omega(\sigma_{\Sigma})\hskip 3.4truecm (4.11)$$
From (4.11), (3.3) we then get 
$$\hskip 2truecm A_{\Sigma} *_{\Omega} A_R = A_{\Sigma}A_R
\hskip 8.2truecm (4.12) $$
and therefore (4.9) reduces to (4.7). Using (4.11), (3.3) we have
$$\int H_{\Sigma}(z_{\Sigma})*_{\Omega}\rho(z_{\Sigma},z_R)\ dz_R=$$
$$=\int d\sigma_{\Sigma}d\sigma_{\Sigma}'\ \tilde H_{\Sigma}(\sigma_{\Sigma})
\tilde \rho(\sigma_{\Sigma}',0) {\Omega(\sigma_{\Sigma})\Omega(\sigma_{\Sigma}')
\over \Omega(\sigma_{\Sigma}+\sigma_{\Sigma}')}e^{i(\sigma_{\Sigma}+
\sigma_{\Sigma}')z_{\Sigma}} e^{\mu(\sigma_{\Sigma}'\land\sigma_{\Sigma})} $$
(in the rest of this section numerical constants depending on $(2\pi) ^{-1}$
are omitted).
Since $\int \rho(z_{\Sigma},z_R)dz_R = \int e^{i\sigma_{\Sigma}'z_{\Sigma}}
\tilde \rho(\sigma_{\Sigma}',0) d\sigma_{\Sigma}'$ we immediately find that
$$\int H_{\Sigma}(z_{\Sigma})*_{\Omega}\rho(z_{\Sigma},z_R)dz_R 
=H_{\Sigma}(z_{\Sigma})*_{\Omega}\int \rho(z_{\Sigma},z_R)dz_R =$$
$$\hskip 2truecm =H_{\Sigma}(z_{\Sigma})*_{\Omega}f_{\Sigma}(z_{\Sigma})\hskip
 8.56truecm  (4.13)$$
Therefore using (4.12), (4.13) and the associativity of the $*_{\Omega}$-product
we obtain
$$P([H_{\Sigma},\rho])=[H_{\Sigma},P{\rho}]$$
hence (4.8a) holds. On the other hand, in the same way using (4.12), (4.10), we
have
$$([H_R,\cdot]P)\rho=[H_R,\rho_R*_\Omega f_{\Sigma}]=f_{\Sigma}[H_R,\rho_R] =0$$
so that the 2nd of (4.8b) holds. Finally with the aid of (3.5)
$$(P[H_R,\cdot])\rho=\rho_R\int[H_R,\rho]\ dz_R=$$
$$=\rho_R\int d\sigma_{\Sigma}d\sigma_Rd\sigma_{\Sigma}'d\sigma_R'dz_R
\biggl( \tilde H_R(\sigma_R)\delta(\sigma_{\Sigma})
\tilde \rho(\sigma_{\Sigma}',\sigma_R')$$
$${\Omega(\sigma_{\Sigma})\Omega(\sigma_R)\Omega(\sigma_{\Sigma}')
\Omega(\sigma_R') \over \Omega(\sigma_{\Sigma}
+\sigma_{\Sigma}')\Omega(\sigma_R+\sigma_R')}
e^{[i(\sigma_{\Sigma}+\sigma_{\Sigma}')z_{\Sigma}+i(\sigma_R+\sigma_R')z_R]}$$
$$2\sinh\mu\{(\eta_{\Sigma}',\eta_R')
\cdot(\xi_{\Sigma},\xi_R)-(\eta_{\Sigma},\eta_R)
\cdot(\xi_{\Sigma}',\xi_R')\}\biggr)=0$$
as simple reductions show. 
\par
The formal adjoint $P^+$ of $P$ can be found as follows: For any 
$A,\rho \in F(\Gamma)$
$$<A,\rho>\equiv\int A^**_{\Omega}\rho\ dz_{\Sigma}dz_R$$
is a bilinear, hermitian form (c.f. [7] p.116-117) and we define
$$<P^+A,\rho>\equiv<A,P{\rho}> \ \ \ \ \ \Leftrightarrow$$
$$\int(P^+A)^**_{\Omega}\rho\ dz_{\Sigma}dz_R = 
\int A^**_{\Omega}(\rho_R*_{\Omega}f_{\Sigma})\ dz_{\Sigma}dz_R=$$
$$\int(\int A^**_{\Omega}\rho_R)\ dz_R)*_{\Omega}f_{\Sigma}dz_{\Sigma}=
\int(\int A^**_{\Omega}\rho_R\ dz_R')*_{\Omega} \rho\  dz_{\Sigma}dz_R$$
Therefore $P^+A=(\int A^**_{\Omega}\rho_R\ dz_R)^*$. Assuming that
$(3.16')$ holds and $\rho_R$ is real we finally have
$$\hskip 2truecm P^+A=\int \rho_R*_{\Omega}A\ dz_R\hskip 8truecm  (4.14) $$
In the same way we define the Liouville operator for the real Hamiltonian $H$,
 by
$$L={i\over 2\mu}[H,\cdot] $$
hence for any $A,\ \rho$, we have $<A,L\rho>=<L^+A,\rho>$. But
$$<A,L\rho>=\int A^**_{\Omega}L\rho\ dz_{\Sigma}dz_R=
\int L(A^**_{\Omega}\rho)\ dz_{\Sigma}dz_R-
\int(LA^*)*_{\Omega}dz_{\Sigma}dz_R $$
By (3.5) the first term is zero hence
$$<A,L\rho>=-{i\over 2\mu}\int[H,A^*]*_{\Omega}\rho\ dz_{\Sigma}dz_R=
{i\over 2\mu}\int[H,A]^**_{\Omega}\rho\ dz_{\Sigma} dz_R$$
where $(3.16')$ and the reality of $H$ have been used. Therefore $L$ is formally
self-adjoint.
\par
Summarizing the above results we may say that {\it if (3.16), (3.17), (4.11)
hold} then the formalism in [9] is valid using the $*_{\Omega}$-product. In
particular the expressions for the generator $\Phi$ (eqs (4.4), (4.22), (4.21)
of [9]) are valid provided that the usual product of functions is everywhere
replaced by the $*_{\Omega}$-product.
\par
In section 1 we remarked that this formalism is valid under the assumption that 
$L_{\Sigma}$ has a point spectrum, a rather severe restriction for 
{\it classical} 
systems. It was also mentioned that in {\it specific} examples in which the 
system ${\Sigma}$ under consideration is approximated by a system 
${\Sigma}_{\omega}$, 
depending on some parameter ${\omega}$ for which $L_{\Sigma_{\omega}}$ has a
point spectrum and for which $L_{\Sigma_{\omega}} \rightarrow L_{\Sigma}$ as
$\omega \rightarrow \omega_0$ say, the corresponding kinetic equation has no 
limit ([16], [21] ch 7). In [11] it has been conjectured that this
may be due to the fact that the perturbation is via a {\it differential} (hence
unbounded in general) operator $L_I$. The present formalism gives the 
possibility to treat the problem more generally: To consider a
$*_{\Omega}$-product depending on two parameters $a,b$ such that the one
measures the discreteness of the spectrum of $L_{\Sigma}$, the other the
deviation of $[\ ,\ ]_{\Omega}$ from the Poisson bracket and take the 
corresponding
limits of the general kinetic equation for the state of $\Sigma$ (c.f. the 
comments on the $*_{\lambda}$ product in section 3 and footnote in this
section).
This will be considered in paper III.
\vskip 2truecm 
\leftline{\bf 5. PHASE-SPACE KINETIC EQUATIONS FOR QUANTUM OPEN}
\leftline{\bf \hskip 0.6truecm SYSTEMS }
\bigskip
In what follows we will apply the formalism developped is section 4 to quantum
open systems weakly coupled to a large reservoir at an equilibrium state
$\hat \rho_R$. To this end we will employ the general formalism developped in 
[9] 
as applied to quantum systems, a brief account of the basic assumptions of which
has been given in the previous section. In fact it is not difficult to see that
by making use of the correspondence rules
$$\hskip 2truecm {\hat A\hat B + \hat B\hat A \over 2} 
\equiv [\hat A,\hat B]_+ \leftrightarrow AB\hskip 6.8truecm (5.1a)$$
$$\hskip 2truecm {1 \over i\hbar}[\hat A ,\hat B] \leftrightarrow 
\{A,B\}\hskip 8.6truecm (5.1b)$$
$$\hskip 2truecm \hat P{\hat \rho} \equiv \hat \rho_R T{r_R} \hat 
\rho \leftrightarrow P\rho = \rho_R \int dx_R\ \ 
\rho(x_{\Sigma},x_R)\hskip 3.6truecm (5.1c)$$
$$\hskip 2truecm \hat P^+ \hat A^+ = 
Tr_R(\hat \rho_R \hat A^+) \leftrightarrow P^+A^* = 
\int dx_R \rho_R A\hskip 3.3truecm (5.1d) $$
$$\hskip 2truecm \hat A^+ \leftrightarrow A^*\hskip 10.4truecm (5.1e)$$
we can obtain the corresponding general kinetic equation for a {\it quantum} 
open system in exactly the same way we did for its classical counterpart, eqs.
 $(4.4), (4.4'), (4.22), (4.22') $ there and under the same assumptions ([9]
section 2, c.f. section 4.2 here). For the sake of brevity we explicit the 
equation for the state $\hat \rho$ of the open system $\Sigma$
\footnote{$^{(7)}$}
{\myfo Notice that other approaches lead to formally identical results (e.g.
[22]-[24], see also below).}.
$${\partial \hat \rho \over \partial t} = - {i \over \hbar} [\hat H_{\Sigma} +
\lambda^2 \hat F, \hat \rho]\hskip 10.7truecm$$
$$\hskip 0.8truecm - {\lambda^2 \over 2\hbar^2} \lim_{T\rightarrow +\infty}
{1 \over 2T} \int_{-T}^{T}dt \int_{-\infty}^{+\infty}ds\ Tr_R 
([\hat H_I(t), [\hat H_I(t+s), \hat \rho_R \hat \rho]])\hskip 1.7truecm (5.2) $$
where
$$\hskip 0.9truecm \hat F = {i \over 2\hbar} 
\lim_{T \rightarrow +\infty} {1 \over 2T}
\int_{-T}^{T} dt \int_0^{\infty} ds\ Tr_R (\hat \rho_R[\hat H_I(t), 
\hat H_I(t+s)])\hskip 1.8truecm (5.3a) $$
$$\hskip 2truecm \hat H_I(t) = e^{i\hat L_0t}\ \ 
\hat H_I \hskip 8.8truecm (5.3b) $$
(cf. (4.6a)).
For a separable interaction\footnote{$^{(8)}$}{\myfo Notice that $\alpha$ is in
general a set of presumably continuous indices.}
$$\hskip 2truecm \hat H_I = \sum_{\alpha} \hat Q_{\alpha}(x_{\Sigma}) 
\hat W_{\alpha}(x_R) =
\sum_{\alpha}\hat Q^+_{\alpha}(x_{\Sigma}) 
\hat W^+_{\alpha}(x_R)\hskip 2.87truecm (5.4) $$
we obtain the quantum analogues of eqs. $(4.22), (4.22')$ of [9] 
(see also [21] section 4.3)
$${\partial \hat \rho \over \partial t} =
 - {i\over \hbar} [\hat H_{\Sigma}, \hat \rho]\hskip 11truecm $$
$$ -{i\lambda ^2 \over 2 \hbar ^2} \biggl[ \sum_{\alpha ,\beta ,\omega}
 {i \bar h^q_{\alpha \beta} (\omega + i0)[\hat V^+_{\alpha}(\omega),
 \hat V_{\beta}(\omega)]
+ i \bar g^q_{\alpha \beta}(\omega + i0) 
[\hat V^+_{\alpha}(\omega), \hat V_{\beta}(\omega)]_+ ), \hat \rho}\biggr]  $$
$$ -{\lambda ^2 \over 2 \hbar} \sum_{\alpha ,\beta ,\omega} 
\biggl(\tilde h^q_{\alpha \beta}(\omega) {1 \over \hbar} 
\biggl[\hat V^+_{\alpha}(\omega),[\hat V_{\beta}(\omega), \hat \rho]\biggr] + 
{\tilde g^q_{\alpha \beta}(\omega) \over \hbar} 
\biggl[\hat V^+_{\alpha}(\omega),[\hat V_{\beta}(\omega), 
\hat \rho]_+\biggr] \biggr) $$
\rightline{(5.5)}
where
$$V_{\alpha}(\omega) \equiv {\cal F}_{\omega}Q_{\alpha},$$
${\cal F}_{\omega}$ being the eigenprojections of $L_{\Sigma}$, and
$$\hskip 2truecm \tilde h^q_{\alpha \beta}(\omega) \equiv 
\int_{-\infty}^{+\infty} ds\ \ e^{i\omega s}
<[\hat W^+_{\alpha}, \hat W_{\beta}(s)]_+>\hskip 3.3truecm (5.6a) $$
$$\hskip 2truecm \tilde g^q_{\alpha \beta}(\omega) 
\equiv \int_{-\infty}^{+\infty} ds\ \ e^{i\omega s}
<[\hat W^+_{\alpha}, \hat W_{\beta}(s)]>\hskip 3.69truecm (5.6b) $$
$$\hskip 2truecm \bar h^q_{\alpha \beta}(\omega + i0) 
\equiv \int_{0}^{+\infty} ds\ \ e^{i\omega s}
<[\hat W^+_{\alpha}, \hat W_{\beta}(s)]_+>\hskip 2.49truecm (5.6c) $$
$$\hskip 2truecm \bar g^q_{\alpha \beta}(\omega + i0) 
\equiv \int_{0}^{+\infty} ds\ \ e^{i\omega s}
<[\hat W^+_{\alpha}, \hat W_{\beta}(s)]>\hskip 2.9truecm (5.6d) $$
$<\hat A>$ denoting $Tr(\hat \rho_R \hat A)$ and 
$\hat W_{\beta}(s) = e^{iL_Rs} \hat W_{\beta}$.

It is clear from (2.13), that an $\Omega$-transformation of (5.5) gives eqs.
$(4.22), (4.22')$ of [9] with the $*_{\Omega}$-product replacing the ordinary
product of functions. This shows explicitly the compatibility of the two methods
to obtain the phase-space expression of a quantum kinetic equation, mentioned
in section 4.

It can be shown, ([21] section 4.3) that expanding the coefficients in (5.5) and
using that by the stationarity of $\hat \rho_R$ 
(i.e. $[\hat H_R, \hat \rho_R]=0$), 
$h_{\alpha \beta}(s) = h^*_{\beta \alpha}(-s)$
where
$$\hskip 2truecm h_{\alpha \beta}(s) 
\equiv <\hat W^+_{\alpha} \hat W_{\beta}(s)>\hskip 7.6truecm (5.7)$$
(5.5) takes the more familiar form (e.g. [22] eq.(III.19), [23] eq.(4.9))
$${\partial \hat \rho \over \partial t} 
= -{i \over \hbar}[\hat H_{\Sigma}, \hat \rho]
+ {i \lambda ^2 \over \hbar} \biggl[{1 \over \hbar} \sum_{\alpha ,\beta ,\omega}
s_{\alpha \beta}(\omega + i0) \hat V^+_{\alpha}(\omega)
\hat V_{\beta}(\omega), \hat \rho\biggr]+\hskip 3truecm$$
$$\hskip 1truecm +{\lambda ^2 \over 2\hbar ^2} 
\sum_{\alpha ,\beta ,\omega} 
\tilde h_{\alpha \beta}(\omega)\biggl(\biggl[\hat V_{\beta}(\omega)
\hat \rho , \hat V^+_{\alpha}(\omega)\biggr] +
\biggl[\hat V_{\beta}(\omega) , \hat \rho 
\hat V^+_{\alpha}(\omega)\biggr]\biggl)\hskip 2truecm (5.8) $$
where
$$\hskip 2truecm \tilde h_{\alpha \beta}(\omega) = 
\int_{-\infty}^{+\infty} ds\ \ e^{i\omega s}
h_{\alpha \beta}(s) ={\tilde h}^*_{\beta \alpha}(\omega)
\hskip 4.3truecm (5.9a) $$
$$\hskip 2truecm s_{\alpha \beta}(\omega + i0) 
\equiv (\Im\bar h (\omega + i0))_{\alpha \beta}
= {\bar h_{\alpha \beta}(\omega + i0) - \bar h_{\beta \alpha}^*(\omega + i0)
 \over 2i }\hskip 0.96truecm (5.9b) $$
$$\hskip 2truecm \bar h_{\alpha \beta}(\omega + i0) = 
\int_0^{+\infty} ds\ e^{i\omega s} h_{\alpha \beta}(s)
\hskip 5.55truecm (5.9c) $$
Using that $\tilde h_{\alpha \beta}(\omega)$ is a nonegative-definite matrix
(see Appendix), equation (5.8) when expanded, can be put in the form
$$ {\partial \hat \rho \over \partial t} = 
-{i \over \hbar} [\hat H_{\Sigma}, \hat \rho] +
+ {i \lambda ^2 \over \hbar}\biggl[{1 \over \hbar}\sum_{\alpha ,\beta ,\omega}
s_{\alpha \beta}(\omega + i0) \hat V_{\alpha}^
+(\omega) \hat V_{\beta}(\omega)\biggr]$$
$$\hskip 2truecm - {\lambda ^2 \over \hbar ^2} 
\sum_{\gamma ,\omega}\biggl(\biggl[\hat U_{\gamma}^
+(\omega) \hat U_{\gamma}(\omega), \hat \rho\biggr]_+ 
- \hat U_{\gamma}(\omega) \hat \rho \hat U^+_{\gamma}(\omega)\biggr)
\hskip 3truecm (5.10) $$
where
$$\hskip 2truecm \hat U_{\gamma}(\omega) 
\equiv \sum_{\beta} k_{\gamma \beta}(\omega) 
\hat V_{\beta}(\omega)\hskip 6.9truecm (5.10'a) $$
$$\hskip 2truecm \tilde h_{\alpha \beta}(\omega) 
= \sum_{\gamma} k^*_{\gamma \alpha}(\omega) k_{\gamma \beta}(\omega)
\hskip 6.6truecm (5.10'b) $$
This is of the form
$$\hskip 2truecm {\partial \hat \rho \over \partial t}
= - {i \over \hbar} [\hat H_{\Sigma} + \lambda ^2 \hat F, \hat \rho] 
- {\lambda ^2 \over \hbar ^2}[\hat R,\hat A]_+ +
{\lambda ^2 \over \hbar ^2} \sum_n \hat A_n \hat \rho \hat A_n^
+\hskip 2truecm (5.10'') $$
where
$$\hat A_n \equiv \hat U_{\gamma}(\omega)\quad\ , \quad\ \hat R 
\equiv \sum_n\hat A_n^+ \hat A_n $$
$$ \hat F = - {1 \over \hbar} \sum_{\alpha ,\beta ,\omega} 
s_{\alpha \beta}(\omega + i0) \hat V_{\alpha}^+(\omega) 
\hat V_{\beta}(\omega) $$
That is, it has the form of the Lindblad generator of a semigroup conserving 
density matrices ([24] Theorem 2). Notice that simple reductions transforms 
(5.8), $(5.10'')$ to the form
$${\partial \hat \rho \over \partial t} = -{i \over \hbar} [\hat H_{\Sigma} + 
\lambda ^2 \hat F, \hat \rho ] - $$
$$-{\lambda ^2 \over 2\hbar ^2} \sum_{n,m} h_{nm}
\biggl({1\over 2} [\hat B^+_n, [\hat B_m, \hat \rho]] 
+ {1\over 2} [\hat B_m, [\hat B_n^+, \hat \rho]]+\hskip 5truecm$$
$$\hskip 3.5truecm + [\hat B^+_n, [\hat B_m, \hat \rho]_+] 
- [\hat B_m, [\hat B_n^+, \hat \rho]_+]\biggl)\hskip 4truecm (5.8')$$
$${\partial \hat \rho \over \partial t} = -{i \over \hbar} [\hat H_{\Sigma} + 
\lambda ^2 \hat F, \hat \rho ] -\hskip 9.5truecm$$
$$- {\lambda ^2 \over 2\hbar ^2} \sum_n 
\biggl({1\over 2} [\hat A^+_n, [\hat A_n, \hat \rho]] 
+ {1\over 2} [ \hat A_n, [\hat A_n^+, \hat \rho]] 
+ [\hat A^+_n, [\hat A_n, \hat \rho]_+] 
- [\hat A_n, [\hat A_n^+, \hat \rho]_+]\biggr)$$
\rightline {(5.11)} 

\ind where $\hat B_n  
\equiv \hat V_{\alpha}(\omega),\ h_{nm} = \tilde h_{\alpha \beta}(\omega) $.

The formalism developped in the previous sections gives us the possibility to 
obtain a phase-space counterpart of the kinetic equation for $\hat \rho$ in any
of the {\it equivalent} forms $(5.2),(5.5),(5.8'), (5.10)$ or $(5.11),$ by using
generalized Wigner transformations (2.11). To this end it is clear that we need 
explicit expressions for the generalized Wigner transform of operators of the 
form
$$[\hat f, \hat g],\ \ \ [\hat f, \hat g]_+,\ \ \ 
\hat f \hat g \hat h,\ \ \ [\hat f, [\hat g,
\hat h]],\ \ \ [\hat f,[\hat g, \hat h]_+],\ \ \ [\hat f, [\hat g, \hat h]]_+ $$
(the last one is needed in the calculation of the adjoint kinetic equation, for
the observables).
\par
These however, are readily obtained from the results of section 3: By (2.13)
$F({\Gamma})$ equipped with the $*_{\Omega}$-product and its associated 
bracket, 
is homomorphic via an $\Omega$-transformation to the algebra of operators.
Therefore the $\Omega$-transform of the above operators are the same, with 
$\hat f$ replaced by $f \equiv \Omega ^{-1}(\hat f)$ etc, and the operator
product replaced by $*_{\Omega}$. In the notation of section (2.1), 
we obtain from
(3.3)
$$[f,g]_{+} = {1\over (2\pi)^2} \int d\sigma d\sigma ' \tilde f(\sigma)
{\tilde g}(\sigma ') {\Omega(\sigma)\Omega(\sigma ')\over \Omega(\sigma
+\sigma ')}
\cosh\mu (\sigma '\land \sigma ) e^{i(\sigma + \sigma ')z}\ \ \ \ (5.12a)$$

$$[f,g] =\hskip 13.7truecm$$
$$\hskip 1.2truecm {1\over (2\pi)^2} \int d\sigma d\sigma ' \tilde f(\sigma)
\tilde g(\sigma ') {\Omega(\sigma)\Omega(\sigma ')\over \Omega(\sigma+\sigma ')}
2\sinh\mu (\sigma '\land \sigma ) e^{i(\sigma +\sigma ')z}
\hskip 1.2truecm (5.12b)$$

$$\widetilde{[f,g]_{+}} = {1\over 2\pi} \int d\sigma ' \tilde f(\sigma ')
\tilde g(\sigma -\sigma ') {\Omega(\sigma ')\Omega(\sigma - \sigma ')
\over \Omega(\sigma)}
\cosh\mu (\sigma \land \sigma ')\hskip 2truecm (5.13a)$$

$$\widetilde {[f,g]} = {1\over 2\pi} \int d\sigma ' \tilde f(\sigma ')
\tilde g(\sigma -\sigma ') {\Omega(\sigma ')\Omega(\sigma - \sigma ')
\over \Omega(\sigma)}
2\sinh\mu (\sigma \land \sigma ')\hskip 2truecm (5.13b)$$
From (3.4), (3.3) an elementary calculation using (2.12), gives
$$f *_{\Omega}(g *_{\Omega} h) =\hskip 12.9truecm$$
$$ {1\over (2\pi)^3}\int d\sigma d\sigma 'd\sigma ''
{\tilde f_w(\sigma)\tilde g_w(\sigma ') \tilde h_w(\sigma '') \over 
\Omega(\sigma +\sigma '+\sigma '')} e^{i(\sigma +\sigma '+\sigma '')z}
e^{\mu(\sigma +\sigma ')\land (\sigma ''+\sigma ')}\hskip 1truecm (5.14) $$
Similarly, using (5.12), (5.13) we get
$$[f,[g,h]] = 
{1\over (2\pi)^3}\int d\sigma d\sigma 'd\sigma ''
{\tilde f_w(\sigma)\tilde g_w(\sigma ') \tilde h_w(\sigma '') 
\over \Omega(\sigma +\sigma '+\sigma '')}\hskip 5.5truecm $$
$$\hskip 1.8truecm e^{i(\sigma +\sigma '+\sigma '')z}
4\sinh\mu(\sigma '\land \sigma '')\sinh\mu(\sigma \land (\sigma '+
\sigma ''))\hskip 2.5truecm  (5.15a)$$
$$[f,[g,h]_+] = 
{1\over (2\pi)^3}\int d\sigma d\sigma 'd\sigma '' 
{\tilde f_w(\sigma)\tilde g_w(\sigma ') \tilde h_w(\sigma '') 
\over \Omega(\sigma +\sigma '+\sigma '')}\hskip 5.3truecm $$
$$\hskip 1.8truecm e^{i(\sigma +\sigma '+\sigma '')z}
(-2)\cosh\mu(\sigma '\land \sigma '')
\sinh\mu(\sigma\land (\sigma '+\sigma '')\hskip 2.1truecm (5.15b) $$
$$[f,[g,h]]_+ = 
{1\over (2\pi)^3}\int d\sigma d\sigma 'd\sigma '' 
{\tilde f_w(\sigma)\tilde g_w(\sigma ') 
\tilde h_w(\sigma '') \over \Omega(\sigma +\sigma '+\sigma '')}
\hskip 5.3truecm $$
$$\hskip 1.8truecm e^{i(\sigma +\sigma '+\sigma '')z}
(-2)\sinh\mu(\sigma '\land \sigma '')\cosh\mu(\sigma \land(\sigma  '
+\sigma ''))\hskip 2.1truecm (5.15c) $$
{\it Assuming that the $\Omega$-transformation is involutive, 
i.e. (3.16) holds,}
eqs (5.15) imply that the phase-space transform of the dissipative part of 
(5.11), i.e. of the sum over $n$, is equal to
$$ {\lambda ^2\over (2\pi)^3 \hbar ^2} \sum_n \int d\sigma d\sigma 'd\sigma ''
\biggl(-\tilde A_n(\sigma)\tilde A_n^*(\sigma ')  
\sinh\mu(\sigma \land (\sigma '+\sigma '')) e^{\mu(\sigma '\land \sigma '')}+$$
$$+\tilde A_n^*(\sigma)\tilde A_n(\sigma ')\sinh\mu(\sigma 
\land (\sigma '+\sigma '')) e^{-\mu(\sigma '\land \sigma '')}\biggr)
\tilde \rho_w(\sigma '')
{e^{i(\sigma +\sigma '+\sigma '')z} \over \Omega(\sigma +\sigma '+\sigma '')}$$
where from now on for the sake of brevity {\it we omit the subscript w from
$\tilde A_n,\tilde B_n$ etc}.
\par
In the second term we make the substitutions $\sigma \rightarrow -\sigma,
\sigma '\rightarrow -\sigma ',\sigma ''\rightarrow -\sigma ''$ and using that
in general ${\tilde A}^*_n(-\sigma) = ({\tilde A}_n(\sigma))^*$
and $\tilde \rho_w(-\sigma '') = ({\tilde \rho}_w(\sigma ''))^*$ since 
$\rho_w$ is real
($\hat \rho = \hat \rho^+$ and proposition 3.4), we find that it is the complex
conjugate of the first term. Hence 
\medskip
\ind $\Omega ^{-1}-transform\ of\ the\ dissipative\ part\ of\ (5.11)\ =$
$$ -{2\lambda ^2 \over (2\pi)^3\hbar ^2}\Re\biggl(
\int d\sigma d\sigma 'd\sigma  ''
{e^{i(\sigma +\sigma '+\sigma '')z} \over \Omega(\sigma 
+\sigma '+\sigma '')}\hskip 7truecm$$
$$\hskip 3truecm \sum_n\tilde A^*_n(\sigma ')\tilde A_n(\sigma) 
\tilde \rho_w(\sigma '')
\sinh\mu(\sigma\land(\sigma '+\sigma ''))
e^{\mu(\sigma '\land\sigma '')}\biggr)\ \ \ (5.16) $$
An exactly similar calculation gives the $\Omega ^{-1}$-transform of $(5.8')$:
$${\partial \rho\over \partial t} = -{i\over \hbar} [H_{\Sigma}, \rho] +
{2\lambda ^2 i\over (2\pi)^3 \hbar ^2} \int d\sigma d\sigma 'd\sigma ''
{e^{i(\sigma +\sigma '+\sigma '')z} \over 
\Omega(\sigma +\sigma '+\sigma '')}\hskip 3truecm$$
$$\hskip 2truecm \bar B(\sigma ',\sigma '') 
\tilde \rho_w(\sigma) e^{\mu(\sigma '\land \sigma '')}
\sinh\mu(\sigma \land (\sigma '+\sigma ''))-\hskip 4truecm $$
$$\hskip 1truecm -{2\lambda ^2 i\over (2\pi)^3 \hbar ^2} \Re 
\int d\sigma d\sigma 'd\sigma ''
{e^{i(\sigma +\sigma '+\sigma '')} \over \Omega(\sigma +\sigma '+
\sigma '')}\hskip 6.6truecm $$
$$\hskip 2truecm B(\sigma ',\sigma ) \tilde \rho_w(\sigma '') 
e^{\mu(\sigma '\land \sigma '')}
\sinh\mu(\sigma \land (\sigma '+\sigma ''))\hskip 3.4truecm (5.17) $$
where
$$\hskip 2truecm  B(\sigma ',\sigma) \equiv \sum_{n,m} \tilde B^*_n(\sigma ') 
h_{nm}
\tilde B_m(\sigma)\hskip 5.71truecm  (5.18a) $$
$$\hskip 2truecm  \bar B(\sigma ',\sigma) \equiv \sum_{n,m} 
\tilde B^*_n(\sigma ') s_{nm}
\tilde B_m(\sigma)\hskip 5.73truecm  (5.18b) $$
$$\hskip 2truecm h_{nm} \equiv \tilde h_{\alpha \beta}(\omega)\ ,\ 
s_{nm}\equiv s_{\alpha \beta}(\omega +i0)\hskip 5.5truecm  (5.18c) $$
Notice that by $\tilde A^*(\sigma) =(\tilde A(-\sigma))^*$, and the 
self-adjointness of the matrices $h_{nm}, s_{nm}$, we get
$$B^*(\sigma ',\sigma)=B(-\sigma,-\sigma ')\ \ \ \ \ \ \ \bar B^*(\sigma ',
\sigma)=\bar B(-\sigma,-\sigma ') $$

\ind {\mybf Remarks}: (i) We do not explicit $[H_{\Sigma}, \rho]$ since
this is a direct application of (5.12b).
\smallskip
(ii) Eq (5.16) is a special case of (5.17) when $h_{nm}=\delta _{nm}$.
\smallskip
(iii) A similar expression, though less symmetric than (5.17), can be obtained
for the phase-space transform of (5.5), having the advantage of showing more
directly the relation of quantum to classical kinetic equations.
\par
From the above discussion we see that a phase-space formulation via 
$\Omega$-transformations of quantum kinetic equations for open systems 
interacting with an equilibrium bath, with a separable interaction, is 
straightforward provided we are able to calculate the {\it Wigner} transforms
of the operators $\hat V_{\alpha}(\omega)$ or 
$\hat A_n$ (c.f. (5.5), $(5.10'')$).
Notice that the correlation matrices $\tilde h_{\alpha \beta}(\omega),\ 
s_{\alpha \beta} (\omega +i0)$ etc are calculated quantum mechanically, 
hence the bath
introduces no additional computational difficulties in this formalism.
\par
Of course as it can be readily seen from (2.13), in the general case, a 
phase-space transformation of (5.2), (5.3) requires the calculation of functions
like
$$\Omega^{-1}(\hat H_I(t)) = e^{itL_I^{\Omega}} H_I $$
where
$$L_I^{\Omega} \equiv {1\over \hbar}[H_I, \cdot],\ \ \ \ \  H_I=\Omega^{-1}(
\hat H_I) $$
not an easy task even in simple examples (see e.g. [7], and the general 
formalism in [8]).Alternatively we may calculate $\Omega^{-1}
(\hat {\cal F}_{\omega} \hat H_I),\ {\cal F}_{\omega}$
being the eigenprojections of $\hat L_0$, (cf. (4.6a)), not an easy 
task either. Moreover, in
this case phase-space quantities of the bath, must be calculated explicitly, 
which
often requires this to be done {\it before} the passage to the thermodynamic
limit of an infinite bath is performed. Hence the present formalism has to be
extended to the case when $\hat q$ or $\hat p$ have discrete spectrum.
Although this 
is possible, and involves no difficulties of principle,
we will not reproduce the calculations here (see the remark at the end of 
section 2) but postpone them till their use in paper III.
\vskip 0.5truecm
\ind {\mybf Acknowledgement}: The motivation for the study of generalized Moyal
structures in section 3 came out of stimulating discussions of Dr. A. Dimakis
and one of us (C.T.).
\vskip 2truecm 
\centerline {\bf APPENDIX }
\bigskip
\par
We must show that for any $\xi_\alpha \in {\cal C},$
$$\sum_{\alpha ,\beta} \tilde h_{\alpha \beta}(\omega) \xi_{\alpha}^{*}
\xi_{\beta} \geq 0$$
Putting $\hat \phi(s) = \sum\limits_{\alpha}\xi_{\alpha} 
\hat W_{\alpha}(s)$, we have 
that
$$ g(s) \equiv <\hat \phi^+\hat \phi(s)> = \sum_{\alpha ,\beta} 
\tilde h_{\alpha \beta}(s) \xi_{\alpha}^{*} \xi_{\beta} $$
But $g(s)$ is a positive-definite function in the sense that is satisfies
$$\sum_{i,j} g(s_i-s_j) z_i z_j^{*} \geq 0\ \ \ \ \ \ \ \ \ \ for\ any\ \ \ 
s_i \in {\cal R},\ \ z_i \in {\cal C}$$
Indeed
$$ \sum_{i,j}g(s_i-s_j)z_iz_j^{*} = \sum_{i,j}<\hat \phi^
+\hat \phi(s_i-s_j)z_iz_j^{*}> =$$
$$ \sum_{i,j}<(z_j\hat \phi)^+e^{-iL_Rs_j}e^{iL_Rs_i}(z_i\hat \phi)>=$$
$$ = \sum_{i,j} <(e^{iL_Rs_j}(z_j\hat \phi))^+(e^{iL_Rs_i}z_i\hat \phi)> =
\sum_{i.j} Tr (\hat \rho_R \hat A_j^+ \hat A_j) \geq 0 $$
since  $\hat \rho_R$  is a positive-defininite operator. Here $ \hat A_j 
\equiv e^{iL_Rs_j}(z_j\hat \phi ) $
and in the 4th equality we used the stationarity of $\hat \rho_R$. Since
$\sum\limits_{\alpha ,\beta} \tilde h_{\alpha \beta}(\omega) \xi_{\alpha}^{*}
\xi_{\beta}$
is the Fourier transform of $g(s)$, the result follows from the Bochner-Wiener-
Khinchine theorem ([27], [17]).
\vfill{\eject}
\centerline{\bf REFERENCES}
\ind 1)(a) I. Prigogine, C. George, F. Henin and L. Rosenfeld, Chemica Scripta 
\ind {\bf 4} (1973) 5

\ind \ \ \ (b) I. Prigogine and A. P. Grecos in:{\it "Problems in the 
Foundations of 
physics"},

\ind \ \ \ G. Toraldo di Francia (ed.) North Holland (1979)

\ind 2) R. Balescu ,{\it "Equilibrium and non-equilibrium statistical 
mechanics"},
\ind \ \ \ Wiley (1975) chs 2, 3, 15-18

\ind \ \ \    R. Balescu {\it "Statistical mechanics of charged particles"}, 
Wiley (1963)

\ind 3) H. Weyl {\it "The theory of groups and quantum mechanics"} 
Dover(1950), ch 4  $\S$ 14

\ind 4) J. E. Moyal Proc. Cambridge Phil Soc {\bf 45} (1949) 99

\ind 5) (a) G. S. Agarwal and E. Wolf, Phys. Rev. D {\bf 2} (1970) 2161, 2187,
 2206

\ind \ \ \    (b) J. R. Klauder and E. C. G. Sudarshan, 
{\it "Fundamentals of quantum optics"},

\ind \ \ \   W. A. Benjamin (1968)

\ind 6) (a) M. Hillery, R. F. O'Conell, M. O. Scully and E P Wigner, Phys.
 Reports

\ind \ \ \ {\bf 106} No3 (1984) 121 

\ind \ \ \   (b) J. G. Kr\"uger and A. Poffyn, Physica {\bf 85A} (1976) 84

\ind 7) F. Bayen, M. Flato, C. Fronsdal, A. Lichnerowicz and D. Sternheimer,
 Ann. Phy- 

\ind \hskip 0.6truecm sics {\bf 110} (1978) 111

\ind 8) M. Gadella, J. M. Gracia-Bondia, L. M. Nieto and J. C. Varilly, 
J. Phys A {\bf 22} (1989) 2709

\ind \ \ \ M. Gadella, L. M. Nieto (1993 preprint) 

\ind 9) A. P. Grecos and C. Tzanakis, Physica A {\bf 151} (1988) 61

\ind 10) C. Tzanakis and A. P. Grecos, Physica A {\bf 149} (1988) 232

\ind 11) C. Tzanakis, Physica A {\bf 179} (1991) 531

\ind 12) E. P. Wigner, Phys Rev {\bf 40} (1932) 749

\ind 13) (a) P. Carruthers and M. M. Nieto, Rev. Mod Phys {\bf 40} (1968) 411

\ind \ \ \     (b) P. G. Newton, Ann. Physics {\bf 124} (1980) 327

\ind \ \ \     (c) M. Moshinsky and T. H. Seligman, Ann. Physics {\bf 114} 
(1978) 243

\ind \ \ \   M. Moshinsky and T. H. Seligman, Ann. Physics {\bf 120} (1979) 402

\ind \ \ \   J. Deenen, M. Moshinsky and T. H. Seligman, Ann. Physics 
{\bf 127} (1980) 458

\ind 14) G. V. Dunne, J.Phys A {\bf 21} (1988) 2321

\ind 15) (a) A. Dimakis and F. M\"uller-Hoissen, Let. Math. Physics {\bf 28} 
(1993) 123

\ind \ \ \     (b) H. Baehr, A. Dimakis and F. M\"uller-Hoissen, J. Phys A 
{\bf 28} (1995) 3197

\ind 16) C. Tzanakis, Physica A {\bf 151} (1988) 90

\ind 17) I. M. Gel'fand and N. Y. Vilenkin, {\it "Generalized functions: 
Applications of

\ind \ \ \ harmonic analysis"}, Academic Press (1964), ch II

\ind 18) P. Fletcher, Phys. Let. B {\bf 248} (1990) 323

\ind 19) W. Pauli, {\it "General principles of Quantum mechanics"}, Springer 
(1980),

\ind \hskip 0.7truecm section 12, p. 100

\ind 20) P. A. M. Dirac, Proc. Roy. Soc. London A {\bf 111} (1926) 281

\ind 21) C. Tzanakis, Ph. D. Thesis, Universit\'e Libre de Bruxelles (1987)

\ind 22) H. Spohn and J. L. Lebowitz, Adv. Chem. Physics {\bf 38} (1978) 109

\ind 23) E. B. Davies, Comm. Math. Physics {\bf 39} (1974) 91

\ind 24) G. Lindblad, Comm. Math. Physics {\bf 48} (1976) 119

\ind 25) W. H. Louisell, {\it "Quantum statistical properties of radiation } 
Wiley Interscience (1973)

\ind 26) I. M. Gel'fand and G. E. Shilov {\it "Generalized functions: 
Properties and 

\ind \ \ \ operators"}, Academic Press (1964)
 
\ind 27) C. Tzanakis, Nuovo Cimento {\bf 108B} (1993) 339

\ind 28) A. P. Grecos, in {\it "Singularities and dynamical systems"} 
S. Pnevmatikos (ed.), 

\ind \ \ \ North-Holland (1983)

\ind 29) A. Frigerio and V. Gorini, J. Math. Phys. {\bf 25} (1984) 1050

\ind 30) A. Dimakis and C. Tzanakis, {\it "Noncommutative geometry and kinetic 

\ind \ \ \ theory"},to appear in J. Phys. A (1996)

\ind 31) C. Tzanakis, {\it "Linear kinetic equations for classical systems: 
Explicit form 

\ind \ \ \ and its mathematical and physical foundations"}, (1995 preprint).}
\end